\documentclass[12pt,american,english]{revtex4}
\usepackage{mathptmx}

\usepackage{courier}

\usepackage[T1]{fontenc}
\usepackage[latin9]{inputenc}
\usepackage{amsmath}
\usepackage{amssymb}
\usepackage{graphicx}

\makeatletter

\@ifundefined{textcolor}{}
{%
 \definecolor{BLACK}{gray}{0}
 \definecolor{WHITE}{gray}{1}
 \definecolor{RED}{rgb}{1,0,0}
 \definecolor{GREEN}{rgb}{0,1,0}
 \definecolor{BLUE}{rgb}{0,0,1}
 \definecolor{CYAN}{cmyk}{1,0,0,0}
 \definecolor{MAGENTA}{cmyk}{0,1,0,0}
 \definecolor{YELLOW}{cmyk}{0,0,1,0}
}

\usepackage{hyperref}

\makeatother

\usepackage{babel}
\begin{document}

\title{Plus Charge Prevalence in Cosmic Rays: Room for Dark Matter in the
Positron Spectrum}

\author{M.A. Malkov$^{1}$, P.H. Diamond$^{1}$ and R.Z. Sagdeev$^{2}$}

\address{$^{1}$CASS and Department of Physics, University of California,
San Diego\\
$^{2}$University of Maryland, College Park}
\begin{abstract}
The unexpected energy spectrum of the positron/electron ratio is interpreted
astrophysically, with a possible exception of the 100-300 GeV range.
The data indicate that this ratio, after a decline between $0.5-8$
GeV, rises steadily with a trend towards saturation at 200-400GeV.
These observations (except for the trend) appear to be in conflict
with the diffusive shock acceleration (DSA) mechanism, operating in
a \emph{single} supernova remnant (SNR) shock. We argue that $e^{+}/e^{-}$
ratio can still be explained by the DSA if positrons are accelerated
in a \emph{subset} of SNR shocks which: (i) propagate in clumpy gas
media, and (ii) are modified by accelerated CR \emph{protons}. The
protons penetrate into the dense gas clumps upstream to produce positrons
and, \emph{charge the clumps positively}. The induced electric field
expels positrons into the upstream plasma where they are shock-accelerated.
Since the shock is modified, these positrons develop a harder spectrum
than that of the CR electrons accelerated in other SNRs. Mixing these
populations explains the increase in the $e^{+}/e^{-}$ ratio at $E>8$
GeV. It decreases at $E<8$ GeV because of a subshock weakening which
also results from the shock modification. Contrary to the expelled
positrons, most of the antiprotons, electrons, and heavier nuclei,
are left unaccelerated inside the clumps. Scenarios for the 100-300
GeV AMS-02 fraction exceeding the model prediction, including, but
not limited to, possible dark matter contribution, are also discussed. 
\end{abstract}
\maketitle

\section{Introduction\label{sec:Introduction}}

Recent measurements of a positron/electron, $e^{+}/\left(e^{-}+e^{+}\right)$,
excess in the $8-300$ GeV range by Pamela, Fermi-LAT and AMS-02 \citep{Pamela_Pos_09,Fermi_Pos_12,AMS_SepElPos14,AMS02_2014}
has added fuel to the hotly contested race for elusive dark matter
(DM) signatures in rapidly improving cosmic ray (CR) data \citep{Hooper09,HooperDM_AMS13,Berezinsky2015JPhCS}.
Indeed, conventional acceleration schemes, even the most promising
of them all, the diffusive shock acceleration (DSA), has not yet suggested
any viable mechanism for the $e^{+}/\left(e^{-}+e^{+}\right)$ anomaly,
free of tension with the antiproton spectra and other secondaries
\citep{Mertsch2009PhRvL,Kachelriess11,CholisHooper2014PhRvD}. 

In addition to the surprising excess at high energies, the $e^{+}/\left(e^{-}+e^{+}\right)$
ratio has a distinct minimum at $\approx8$GeV which is not easier
to explain making a minimum of assumptions. Both features appear at
odds with the \emph{single source} DSA operation, which predicts similar
rigidity ($R=$momentum/charge) spectra for all primary species. Moreover,
there are also other well documented exceptions, namely the He$^{++}/p$
and $C/p$ ratios that both show a $\sim R^{0.1}$ growth, also seemingly
inconsistent with the DSA \citep{Adriani11,AMS02He2015PhRvL,AMS02andBessPtoHe2015}.
Less pronounced than $e^{+}/\left(e^{-}+e^{+}\right)$, but not less
astonishing at first glance, these anomalies can be explained by the
difference in charge to mass ratio \citep{MDSPamela12}. Other scenarios
are possible but require additional assumptions, such as inhomogeneity
of the SNR environment \citep{Ohira11,Ohira15,Drury12} or multiple
sources with adjusted spectral indices (see \citep{Serpico15,Tomassetti2015ApJ,Ohira15}
for a recent discussion). In fact, the mass to charge based explanation
of the $\approx0.1$ difference in rigidity indices has been given
only for the He/$p$ spectrum, while the C/$p$ was measured with
sufficient accuracy only recently \citep{AMS02He2015PhRvL} and turned
out to be identical to the He/$p$ rigidity spectrum. Thus, the mechanism
suggested by \citep{MDSPamela12} predicted the $C/p$ spectrum since
He and C have the same mass to charge ratio. If this injection mechanism
is correct, the latest AMS-02 data speak against a direct carbon acceleration
from grains \citep{Meyer97}.

The mass to charge selectivity of the DSA which works for He/$p$
and C/$p$ does not apply to the $e^{+}/e^{-}$ fraction. It, therefore,
seems logical to look for a possible \emph{charge-sign} dependence
of the SNR-DSA production of CRs, including the $e^{+}/\left(e^{-}+e^{+}\right)$
anomaly. We call it 'anomaly' rather than 'excess' (also encountered
in the literature) since the ratio rises with the particle energy
only at $E>8$GeV. Below this energy, it \emph{declines, }thus creating
a \emph{deficit}. The decline, the rise and the clear minimum between
them (at $8$ GeV) are all pivotal to the mechanism proposed here.
These aspects are intrinsic to a \emph{single-source} mechanism proposed,
revealing unique characteristics of the accelerator. By contrast,
assuming two or more independent positron contributions to the spectrum
(as, e.g., in refs.\citep{ErlykWolf2013APh,Mertsch2014PhRvD,Cowsik2014ApJ}),
one fits the nonmonotonic positron fraction, but with no constraints
on the underlying acceleration mechanisms. The position of the minimum
in the positron fraction is then coincidental, and the fit does not
add credibility to the model predictions for the higher energy data
points yet to come. We will return to this point in the Discussion
section.

A vast majority of conventional scenarios for the $e^{+}/\left(e^{-}+e^{+}\right)$
excess (including the present one) invoke secondary positrons. They
are produced by galactic CR protons colliding with an ambient gas
near an SNR accelerator, e.g. \citep{Fujita2009PhRvD}, elsewhere
in the galaxy, e.g.,\citep{Waxman2013PhRvL,Cowsik2014ApJ}, or are
immediately involved in the SNR shock acceleration, \citep{Blasi2009PhRvL,Mertsch2014PhRvD,CholisHooper2014PhRvD}.
Some of these scenarios face the unmatched antiprotons and other secondaries
in the data, as discussed, e.g., in \citep{Kachelriess11,VladimirMoskPamela11,CholisHooper2014PhRvD}.
Improvements along these lines have recently been achieved by using
Monte Carlo $pp$ collision event generators, e.g. \citep{Kohri2016PTEP}.
However, improved cross sections of $pp$ collisions do not shed light
on the \emph{physics} of $e^{+}/\left(e^{-}+e^{+}\right)$ anomaly,
particularly the minimum at 8 GeV. This spectrum complexity hints
at richer physics than a mere production of secondary $e^{+}$ and
$\bar{p}$ power-law spectra from the primary CR power-law. 

We propose and investigate the idea that the physics of the $e^{+}/\left(e^{-}+e^{+}\right)$
fraction and, by implication, that of the $\bar{p}/p$ unmatched fraction,
is in the charge-sign asymmetry of particle acceleration. The subsequent
particle propagation through the galaxy or multiple accelerators plays
no significant role in the phenomenon, as they act equally on all
species. This proposition is particularly consistent with a scenario
wherein almost all the positrons contributing to the observed $e^{+}/\left(e^{-}+e^{+}\right)$
ratio are produced in a single SNR of a particular kind described
further in the paper. Electrons in the $e^{+}/\left(e^{-}+e^{+}\right)$
fraction may in part originate from an ensemble of other remnants.
However, the single-source explanation for the $e^{+}/\left(e^{-}+e^{+}\right)$
anomaly is \emph{generic} to all SNRs of the kind and thus is equally
consistent with its multi-source origin. From the Occam's razor perspective,
this mechanism is preferred over those requiring \emph{different }types
of sources, to state the obvious. 

A striking exception to the proposed scenario is the 100-300 GeV range
where the current AMS-02 points significantly exceed our model predictions.
This region then requires an independent source atop of the SNR contribution,
that can be of a dark matter annihilation/decay or pulsar origin.
Further model improvements are planned to see if simplifications made
in its current version are responsible for the difference, but it
appears to be unlikely.

The proposed mechanism relies on the following two aspects of the
DSA. The first one is the injection process whereby particles become
supra-thermal and may then cross and re-cross the shock front, thus
gaining more energy. The proposed injection mechanism is charge-sign
asymmetric. It differs from the conventional DSA in which the injection
efficiency primarily depends on the mass-to-charge ratio but not so
much on the sign of the charge (see, however, the Discussion section).
We will argue that the charge-sign dependence of injection arises
when the shock propagates into an interstellar medium (ISM) containing
clumps of dense molecular gas (MC, for short).

The second aspect of the proposed mechanism concerns the phenomenon
of nonlinear shock modification which is known to make the spectrum
of low-energy particles steeper and that of the high-energy particles
flatter than the canonical $p^{-4}$ spectrum produced by strong but
unmodified shocks. Consequently, in the modified shocks, a point $p=p_{4}$
exists in the particle energy spectrum where the index is equal to
four. Assuming that the bulk of galactic CR electrons are accelerated
in conventional shocks, thus having $p^{-4}$ source spectra, the
ratio of the modified positron spectrum to unmodified electron spectrum
will show the required nonmonotonic behavior with a minimum at $p=p_{4}$.
In a customary $p^{4}f$$\left(p\right)$ normalization, the individual
positron spectrum is, therefore, the same as that of the $e^{+}/e^{-}$
ratio. Therefore, it conicides with the \emph{proton }spectrum, provided
all the species are \emph{relativistic}. An analytic solution places
the \emph{proton} $p^{4}f$$\left(p\right)$ minimum at $\lesssim10$
GeV/c (see Fig.5 in Ref. \citep{MDru01}), depending weakly on the
shock Mach number, $M$, proton maximum energy, $E_{{\rm max}}$,
and their injection rate. However, $M\gtrsim10$ and $E_{\max}\gtrsim1$
TeV conditions are required, along with some minimum proton injection,
for the solution to transition into a strongly nonlinear regime (often
called efficient acceleration). Although the minimum in the spectrum
looks encouraging for explaining the nonmonotonic $e^{+}/e^{-}$ ratio,
it was obtained for protons and needs to be reconsidered for positrons
in the 1-10 GeV/c momentum range. The reason for that is a different
momentum dependence of positron and proton diffusivity (positrons
enter a relativistic transport regime at much lower energy than protons). 

Once an SNR shock is strongly modified, MCs in its precursor will
survive the sub-shock UV and X- radiation, severely diminished in
such shocks. At the same time, shock-accelerated CR protons illuminate
the MC well before the subshock encounter. These CRs generate positrons
(along with other secondaries) in the MC interior by colliding with
the dense gas material. The CR protons also charge the MC \emph{positively;}
as a result, many positively charged particles abandon the MC, while
negatively charged particles remain inside. Being charged by the shock-accelerated
protons, the MC thus acquires a positive potential which creates a
charge-sign asymmetry for the subsequent particle injection into the
DSA.

Plasmas are intolerant to external charges and immediately restore
charge neutrality. Nevertheless, a large and dense MC needs to build
up a strong electric field to restore charge neutrality. Due to the
high rigidity of CRs, their density in the MC interior increases almost
simultaneously with that in the exterior, as a CR-loaded shock approaches
the MC. However, by contrast with a strongly ionized exterior, where
the plasma resistivity is negligible, the electron-ion, ion-neutral
(and, in the case of very dense clouds, also electron-neutral) collisions
inside the MC, provide significanty resistivity to the neutralizing
electric current. Therefore, a strong macroscopic electric field is
generated in response to the CR penetration. This field expels the
secondary positrons most efficiently as the lightest positively charged
species \textendash{} although it also shields the MC from low-energy
CR protons.

The mechanism outlined above implies that negatively charged primaries
and secondaries have much better chances to stay in an MC than positively
charged particles. When the subshock eventually reaches the MC, the
subshock engulfs it, e.g., \citep{Inoue12,Draine2010BookISM}. What
was inside of the MC, is transferred downstream unprocessed by the
subshock. Therefore, the negatively charged particles in its interior
largely evade acceleration. This charge-sign asymmetry of particle
injection into the DSA explains why there is no $\bar{p}/p$ excess,
similar to that of $e^{+}/e^{-}$ .

It follows that the positron spectrum results from several interwoven
processes. We will consider them separately and study their linkage.
The remainder of the paper is organized as follows. Sec.\ref{sec:InteractionOfCRwithMC}
deals with a spatial distribution of CR in a shock precursor, their
propagation inside of an MC and electrodynamic processes that the
CR induce there. In Sec.\ref{sec:Spectrum-of-shock-accelerated} we
discuss and estimate the distribution of secondary positrons as they
come out of the MC and become subject to the DSA. The spectrum of
accelerated positrons is calculated from low to high energies, and
the nature of the 8 GeV minimum is elucidated. We briefly discuss
some of the alternative explanations of the $e^{+}/\left(e^{-}+e^{+}\right)$
excess in Sec.\ref{sec:Discussion}, followed by the Conclusion section,
Sec.\ref{sec:Conclusions}.

\section{Interaction of shock-accelerated \protect \\
protons with MC\label{sec:InteractionOfCRwithMC}}

An MC illumination by shock-accelerated CR protons before the shock
arrival is crucial for the mechanism of positron injection into the
DSA. The protons begin interacting with the MC when its distance to
the subshock shortens to the CR diffusion length $\sim\kappa/u_{1}$,
Fig.\ref{fig:ShockMCinteraction}. Here $\kappa$ is the CR diffusion
coefficient and $u_{1}$ is the shock velocity. Generally, $\kappa$
depends on the CR momentum, e.g., in a Bohm limit, $\kappa\simeq cr_{g}\left(p\right)/3$,
where $r_{g}$ is the CR gyro-radius. As $\kappa$ grows with $p$,
higher energy protons reach the MC earlier. To understand the electrodynamic
response of the MC to the penetrating CRs, we need to know their number
density depending on the subshock distance. This subject is addressed
in the next subsection. 

\subsection{CR spatial profile in unshocked plasma\label{subsec:CR-spatial-profile}}

The simplest assumption to start with is that CRs penetrate freely
into an MC with an implication that their number density, $n_{CR}$,
inside the MC depends only on the distance to the subshock, $x_{MC}$.
The assumption is reasonable for high energy CRs and small MCs, as
the CRs penetrate the MC more easily and omnidirectionally in this
case. It is generally not valid for large MCs \citep{MDS_APS12},
which we discuss in Appendix \ref{sec:AppCRinsideMC}. We also show
there that, in a wide range of conditions, the CR number density can
be approximated by 

\begin{equation}
n_{CR}\left(x\right)=\frac{x_{0}n_{CR}^{0}}{x_{0}+x_{MC}}\label{eq:NcrOfx}
\end{equation}
Here \foreignlanguage{english}{$n_{CR}^{0}$} is the CR density at
the subshock ($x_{MC}=0$) and $x_{0}$ weakly depends on the CR momentum
distribution. Note that the last expression is virtually independent
of the degree of shock modification. However, in modified shocks the
flow velocity gradually decreases from its far upstream value $u_{1}$
to $u_{0}$ ahead of the subshock, where it drops abruptly to $u_{2}<u_{0}$
downstream, Fig.\ref{fig:Flow-profile-near}. The total shock compression
ratio $r=u_{1}/u_{2}$ depends on the shock Mach number and the CR
pressure \citep{MDru01} and may be much higher than the typical value
of four. The subshock compression ratio, $r_{s}=u_{0}/u_{2}$ can,
in turn, be significantly lower than four. 

While the plasma slows down towards the subshock, an MC proceeds at
a higher speed, Fig.\ref{fig:Flow-profile-near} because the CR pressure
and the ram pressure of the plasma are insufficient to slow down dense
clouds considerably. Thus, the MCs encounters a supersonic headwind
(for $u_{1}-u_{0}>C_{s}$, the sound speed) and a bow-shock must form
on the shock side of the MC. However, we will not discuss this further
in the paper. Instead, we focus on the plasma processes in the MC
interior initiated by penetrating CR protons. In what follows, we
assume them to have enough energy to cross the possible bow-shock
and the MC-plasma interface unimpededly.

As a rule, plasmas respond promptly to external charges and readily
restore its neutrality due to high electric conductivity. But when
energetic protons penetrate into a more resistive plasma inside an
MC, a stronger electric field needs to build up to neutralize the
charge, and there are several neutralization scenarios to consider.
First, because of the electric field, the MC plasma may suck in external
thermal electrons. Again, the MC plasma is electrically resistive
due to a high neutral density and low temperature. Therefore, the
efficiency of this neutralization is limited, and the electrostatic
potential inside the MC may grow considerably to sustain the charge
neutrality, possibly up to an electric breakdown of the MC neutral
gas with significant pair production \citep{GurevichPairs00}. Another
limitation comes from the total flux of thermal electrons entering
the MC. It cannot significantly exceed the value $\sim V_{Te}n_{0}S$,
where $n_{0}$ and $V_{Te}$ are the electron density and thermal
velocity, while $S$ is the effective MC cross section across the
magnetic field. (We neglect the cross-field particle transport here
and below, including that of the CRs). 

To facilitate our discussion of the MC neutralization, we introduce
a charge budget parameter, $\eta$, as a ratio of the MC charging
rate by CR protons to its neutralization rate by inflowing extraneous
electrons and outflowing MC ions:

\begin{equation}
\eta=\frac{\dot{n}_{{\rm CR}}L_{{\rm MC}}}{V_{Te}n_{0}+V_{i}n_{i}}\sim\frac{L_{{\rm MC}}}{L_{{\rm CR}}}\cdot\frac{u_{1}n_{{\rm CR}}}{V_{Te}n_{0}+V_{i}n_{i}}\label{eq:etaParam}
\end{equation}
Here $\dot{n}_{CR}=dn_{CR}/dt$ is the CR charging rate, $n_{i}$
and $V_{i}$ are the density and velocity of the ions at the MC boundaries,
$x=\pm a$. We count the $x$- coordinate from the \foreignlanguage{american}{center}
of the MC along the field line, Fig.\ref{fig:ShockMCinteraction}.
The characteristic time of CR increase in the MC (also the CR maximum
acceleration time) is $t_{a}\sim L_{CR}/u_{1}$, where $L_{CR}\sim\kappa\left(p_{{\rm max}}\right)/u_{1}$
is the CR precursor scale-height and $u_{1}$ is the shock velocity.
At least initially, the charge neutralizing current is carried by
the thermal electrons (first term in the denominator). However, it
does not increase in response to the growing electric field as this
flux is fixed by the ambient plasma conditions, not affected by the
MC. By contrast, the ion contribution to the CR neutralization (second
term in the denominator in \foreignlanguage{american}{eq}.{[}\ref{eq:etaParam}{]})
is dynamic and becomes more important when the electric field accelerates
the outflowing ions. At the same time, the resulting ion depletion
inside the MC may eventually diminish neutralization. It follows that
the parameter $\eta$, although small in general, may grow significantly,
especially in strong SNR shocks where $u_{1}\gg V_{Te}$. The resulting
electric field $E\left(x\right)$ clearly depends on $\eta$ so that
we include the above aspects in an equations for $E$ in the next
subsection. 

The time dependence of $n_{CR}\left(t\right)$ regulates an MC charging.
We substitute $n_{CR}\left(x\right)$ from eq.(\ref{eq:NcrOfx}) and
assume that the MC propagates ballistically through the shock precursor,
that is we can write $x_{MC}=-u_{1}t$, where $-\infty<t\leq0$ and
the subshock reaches the MC at $t=0$. Therefore, $n_{CR}$ grows
in time as
\begin{equation}
n_{CR}\left(t\right)=n_{CR}^{0}x_{0}/\left(x_{0}-u_{1}t\right)\label{eq:NcrOft}
\end{equation}
As $n_{CR}$ grows rapidly when the MC approaches the subshock (the
growth stops when $n_{CR}$ reaches $n_{CR}\left(t=0\right)=n_{CR}^{0}$),
the following reaction from the MC is expected. First, the increase
in $\eta$ may slow down, as the electric field ejects more ions from
the MC. But, when many of them are expelled and the ion flux $V_{i}n_{i}$
cannot balance the continuing $n_{CR}$ increase, the electric field
inside the MC may exceed the ionization threshold. As a result, the
$n_{i}$ will increase, thus limiting $\eta$, or even an electric
breakdown of the gas becomes possible, as mentioned earlier. We defer
this issue to future work and consider in the next subsection the
build-up of an electrostatic potential inside the MC, as the latter
is charged by penetrating CRs with neglected ionization and recombination. 

\subsection{Electrodynamics of CR-MC interaction\label{subsec:Electrodynamics-of-CR-MC}}

\begin{figure}
\includegraphics[scale=0.55]{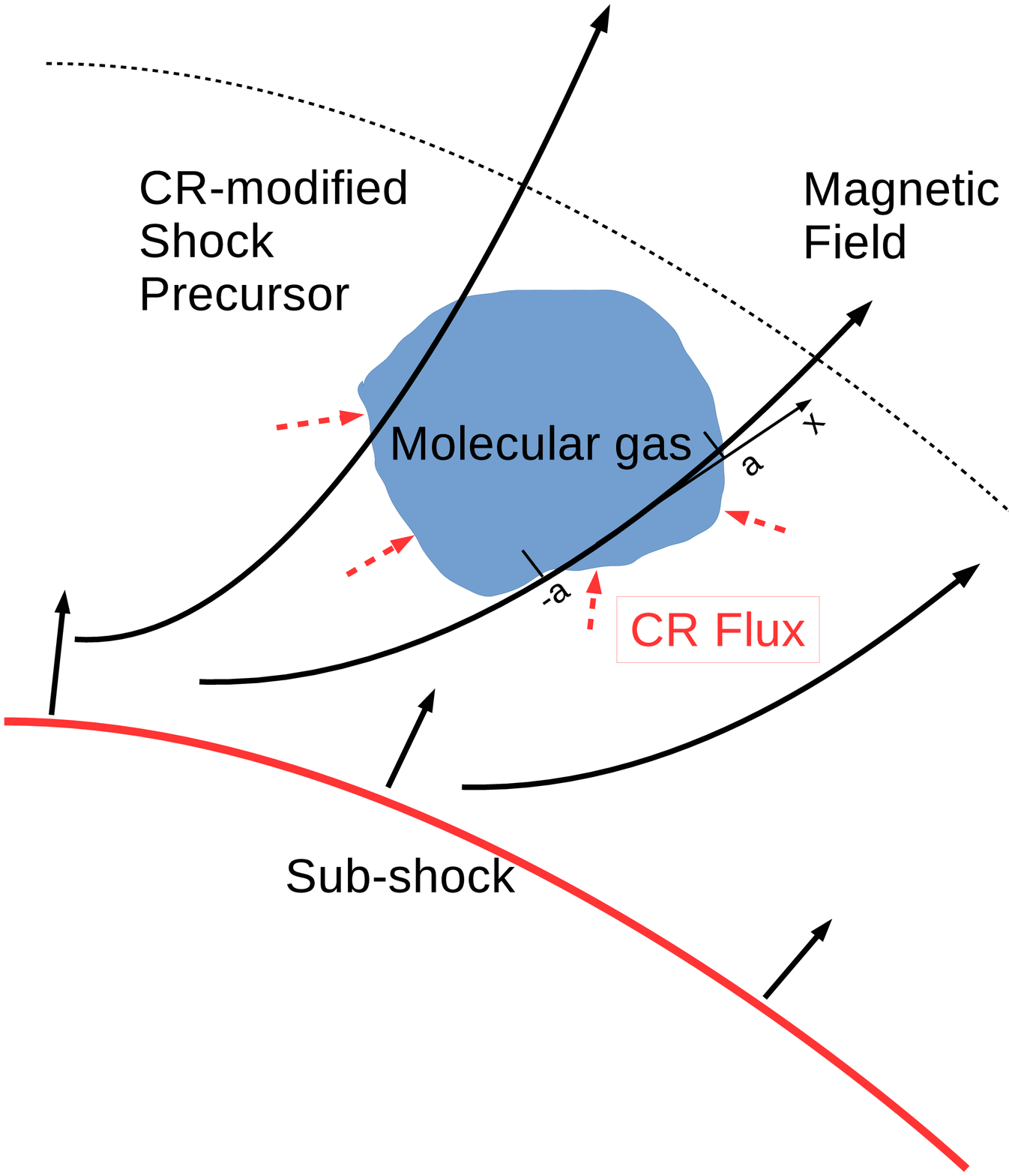}

\caption{SNR shock propagating into ISM with MC upstream.\label{fig:ShockMCinteraction}}

\end{figure}

\begin{figure}
\includegraphics[bb=0bp 220bp 612bp 700bp,clip,scale=0.6]{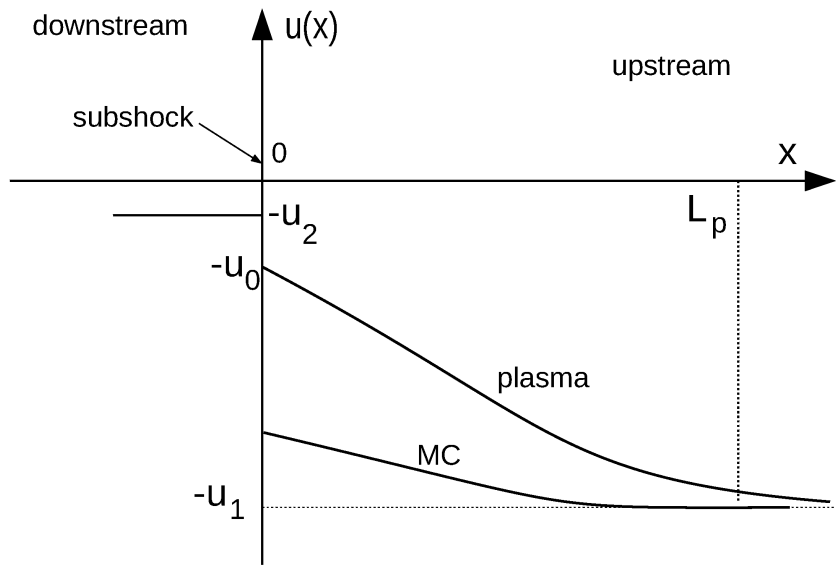}

\caption{Flow profile near a CR modified shock. The line 'MC' shows an MC trajectory
on the phase plane. The speed of MC is considerably higher than that
of the flow because of its inertia, resulting in a weaker slow down
by the CR pressure than that for the main plasma. The drag from the
plasma is also assumed to be not sufficient to slow down the MC significantly.
\label{fig:Flow-profile-near}}
\end{figure}

For describing an MC response to penetrating CR protons, we use two-fluid
equations for electrons and ions that move along the $x-$axis (magnetic
field direction, Fig.\ref{fig:ShockMCinteraction}) in the MC interior:

\begin{eqnarray*}
\frac{dV_{i}}{dt} & = & \frac{e}{m_{i}}E\left(x,t\right)-\nu_{in}V_{i}\\
\frac{dV_{e}}{dt} & = & -\frac{e}{m_{e}}E-\nu_{ei}\left(V_{e}-V_{i}\right)\\
\frac{\partial n_{e,i}}{\partial t} & = & -\frac{\partial}{\partial x}n_{e,i}V_{e,i}\\
n_{e} & = & n_{i}+n_{CR}
\end{eqnarray*}
where 

\[
\frac{d}{dt}\equiv\frac{\partial}{\partial t}+V_{i}\frac{\partial}{\partial x}
\]
Here $V_{i,e}$ and $n_{i,e}$ are the mass velocities and number
densities of electron and ion fluids, $E=-\partial\phi/\partial x$
is the electric field, $\nu_{in}$ and $\nu_{ei}$ are the ion-neutral
and electron-ion collision frequencies. The last equation is the usual
quasi-neutrality condition replacing the Poisson equation because
$L_{MC}$ exceeds the Debye radius by many orders of magnitude. On
comparing the first two equations, we neglect the electron inertia
term in the second equation, after which it suggests eliminating the
electron velocity $V_{e}$ altogether. Furthermore, by taking the
difference between the continuity equations for electrons and ions,
and introducing the CR column density inside the MC, $N_{CR}$, by
the relation $n_{CR}=\partial N_{CR}/\partial x$, the above system
of five equations may be manipulated into the following two equations:

\begin{eqnarray}
\frac{dV_{i}}{dt} & = & -\nu_{in}V_{i}+\frac{m_{e}}{m_{i}}\left(\dot{N}_{CR}+n_{CR}V_{i}\right)\frac{\nu_{ei}}{n_{CR}+n_{i}}\label{eq:dVidt2}\\
\frac{dn_{i}}{dt} & = & -n_{i}\frac{\partial V_{i}}{\partial x}\label{eq:nicont}
\end{eqnarray}
The dot over $N_{CR}$ stands for a time derivative. The second term
on the r.h.s. of eq.(\ref{eq:dVidt2}) is proportional to the electric
field, 

\begin{equation}
E\left(x,t\right)=\frac{m_{e}}{e}\nu_{ei}^{\prime}\frac{n_{CR}n_{i}}{n_{CR}+n_{i}}\left(\frac{\dot{n}_{CR}}{n_{CR}}x+V_{i}\right),\label{eq:Eofxt}
\end{equation}
where we have ignored variations of CR density inside the MC and used
the linear approximation for $N_{CR}\approx n_{CR}x$, along with
a symmetry requirement, $E=V_{i}=0$ at $x=0$ (center of MC). We
have also separated the ion density from collision frequency $\nu_{ei}$,
by introducing the following parameter

\[
\nu_{ei}^{\prime}=\frac{4}{3}\sqrt{2\pi}\frac{e^{4}}{\sqrt{m}T_{e}^{3/2}}\Lambda=\nu_{ei}/n_{i}
\]
where $\Lambda\sim10$ is a Coulomb logarithm. The ion-neutral collision
frequency can be written as follows

\[
\nu_{in}=\frac{8\sqrt{2}}{3\sqrt{\pi}}\sigma_{in}n_{n}\sqrt{\frac{m_{i}T}{m_{n}\left(m_{n}+m_{i}\right)}}
\]
with the ion-neutral collision cross section $\sigma_{in}\approx5\cdot10^{-14}cm^{2}.$
We neglect the electron-neutral collisions, since $\sigma_{en}\approx4\cdot10^{-15}$.
These and other parameters, pertinent to MCs, are summarized, e.g.,
in \citep{Dogiel1987MNRAS}. Using the above approximation of coordinate
independent $n_{CR}$ ($N_{CR}\approx n_{CR}x$), we will convert
the system given by eqs.(\ref{eq:dVidt2}-\ref{eq:nicont}) into a
system of two ordinary differential equations but first, we introduce
some dimensionless variables. It might appear natural to measure time
in precursor crossing times, $t_{a}=L_{CR}/u_{1}=\kappa\left(p_{{\rm max}}\right)/u_{1}^{2}$,
Fig.\ref{fig:Flow-profile-near}. However, since our focus here is
on processes occurring inside the MC, as it traverses the shock precursor,
the MC travel time $t_{a}$ is not the best time unit. Indeed, the
main driver of these processes is the changing $n_{CR}\left(t\right)$
which is nearly scale free, eq.(\ref{eq:NcrOft}). It is, therefore,
more convenient to choose $\nu_{in}^{-1}$ for the time unit. Denoting
by $2a$ the length of a given field line, to which eqs.(\ref{eq:dVidt2}-\ref{eq:nicont})
refer inside the MC (Fig.\ref{fig:ShockMCinteraction}), we use the
following scales of the remaining variables:

\[
\frac{x}{a}\to x,\;\;\nu_{in}t\to t,\;\;\;\frac{V_{i}}{\nu_{in}a}\to V_{i},\;\;\;\frac{n_{i}}{n_{0}}\to n_{i},\;\;\;\frac{n_{CR}}{n_{0}}\to n_{CR}
\]
where $n_{0}$ is the initial ion density. Eqs.(\ref{eq:dVidt2}-\ref{eq:nicont})
rewrite then as follows:

\begin{eqnarray}
\frac{\partial V_{i}}{\partial t} & = & -V_{i}\frac{\partial V_{i}}{\partial x}-V_{i}+\nu_{e}\frac{n_{CR}n_{i}}{n_{CR}+n_{i}}\left(\frac{\dot{n}_{CR}}{n_{CR}}x+V_{i}\right)\label{eq:dVdtResc}\\
\frac{\partial n_{i}}{\partial t} & = & -n_{i}\frac{\partial V_{i}}{\partial x}-V_{i}\frac{\partial n_{i}}{\partial x}\label{eq:dndtResc}
\end{eqnarray}
where we introduced the following collision parameter

\[
\nu_{e}=\frac{m_{e}}{m_{i}}\frac{\nu_{ei}^{\prime}n_{0}}{\nu_{in}}.
\]

We need to solve eqs.(\ref{eq:dVdtResc}) and (\ref{eq:dndtResc})
in the domain $-1<x<1$. For $a\ll L_{CR}$, or roughly also for $a\lesssim L_{CR}$
\emph{and} a quasiperpendicular shock geometry, the following symmetry
conditions are suggestive, $\mbox{\ensuremath{n_{i}\left(-x\right)=n_{i}\left(x\right),} }V\left(-x\right)=-V\left(x\right)$
(see also Appendix \ref{sec:AppCRinsideMC}). They require the following
boundary conditions at $x=0$: $\partial n_{i}/\partial x=0,\;\;V=0$.
Note that the shock geometry becomes progressively quasiperpendicular
towards the subshock by virtue of compressed magnetic field component
in the shock plane, Fig.\ref{fig:ShockMCinteraction}. The electric
field $E$ will have the same symmetry properties as $V.$ With the
above boundary conditions, eqs.(\ref{eq:dVdtResc}- \ref{eq:dndtResc})
admit the following simple form of solution

\begin{eqnarray}
n_{i} & = & n_{i}\left(t\right)\label{eq:niOft}\\
V & = & \psi\left(t\right)x\label{eq:Voft}
\end{eqnarray}
so that eqs.(\ref{eq:dVdtResc}-\ref{eq:dndtResc}) reduce to an ODE
system:

\begin{eqnarray}
\frac{d\psi}{dt} & = & -\psi^{2}-\nu_{i}\psi+\frac{\nu_{e}n_{i}}{n_{CR}+n_{i}}\left(\dot{n}_{CR}+n_{CR}\psi\right)\label{eq:dpsidt}\\
\frac{dn_{i}}{dt} & = & -\psi n_{i}\label{eq:dnidt}
\end{eqnarray}
An assumption $n_{CR}\ll n_{i}$ greatly simplifies the first equation
of this system and allows us to solve it independently of the second
one. The assumption remains plausible during an initial phase of the
MC-CR interaction, but it may be violated at later times when $n_{CR}$
increases while $n_{i}$ decreases because of the ion outflow. When
the condition holds up, the solution of the second equation can be
readily obtained in terms of $\psi,$ while the first equation makes
a Riccati equation for $\psi$. After this equation is solved we will
constrain the problem parameters to ensure the condition $n_{CR}\ll n_{i}$.

The solution to eq. (\ref{eq:dpsidt}) is obtained in Appendix \ref{sec:Appendix:-electrodynamics-inside}
, and the transition from the PDE to ODE is justified by a direct
numerical integration of the original PDE system, given by eqs.(\ref{eq:dVdtResc}-\ref{eq:dndtResc}).
We can write the solution for $\psi$ as follows

\begin{equation}
\psi\left(\tau,\alpha\right)=\frac{\alpha}{\tau}-1+\frac{\tau{}^{\alpha}e^{-\tau}}{\Gamma\left(\alpha+1,\tau\right)}\label{eq:psiOftauandalpha}
\end{equation}
where $\Gamma$ is an incomplete gamma-function,

\[
\Gamma\left(\alpha+1,\tau\right)\equiv\int_{\tau}^{\infty}t^{\alpha}e^{-t}dt
\]
and $\tau=t_{0}-t$. The dimensionless parameters $\alpha$ and $t_{0}$
are defined in eq.(\ref{eq:alphaAndt0}), $t_{0}=\nu_{in}a/u_{1}$,
$\alpha=\left(m_{e}/m_{i}\right)\left(\nu_{ei}a/u_{1}\right)\left(n_{CR}^{0}/n_{i}\right)$.
The density depletion of the MC ions can now be obtained from eq.(\ref{eq:dnidt}):

\begin{equation}
\frac{n_{i}\left(t=0\right)}{n_{i}\left(t=-\infty\right)}=\left[\int_{t_{0}}^{\infty}\left(\tau/t_{0}\right)^{\alpha}e^{t_{0}-\tau}d\tau\right]^{-1}\label{eq:zeta}
\end{equation}
where $t=-\infty,0$ refer to the far upstream ion density, $n_{i}$,
and its value when the subshock intersects the MC. Whether the above
ratio can be considerably smaller than unity, thus possibly violating
the assumption $n_{CR}\ll n_{i}$, depends on the parameter $\alpha/t_{0}$.
This is because a saddle point on the phase integral in eq.(\ref{eq:zeta})
is on the integration path for $\alpha>t_{0}$, thus making a large
contribution to the integral. So, one can estimate the integral as
follows

\[
\int_{t_{0}}^{\infty}\left(\tau/t_{0}\right)^{\alpha}e^{t_{0}-\tau}d\tau=\begin{cases}
1+\alpha/t_{0}, & \alpha/t_{0}<1\\
\sqrt{2\pi\alpha}\left(\alpha/t_{0}\right)^{\alpha}e^{t_{0}-\alpha}, & \alpha/t_{0}>1
\end{cases}
\]

To reconstruct possible scenarios of MC neutralization, we take a
closer look at the parameter $\alpha/t_{0}$ in eq.(\ref{eq:alphaAndt0}).
Assuming for simplicity that $T_{e}\sim T_{i}$ inside the clump,
the following estimate can be obtained:

\[
\frac{\alpha}{t_{0}}\sim\left(\frac{1eV}{T_{e}}\right)^{2}\frac{n_{CR}^{0}}{n_{n}}\sqrt{\frac{m_{n}}{m_{i}}\left(\frac{m_{n}}{m_{i}}+1\right)\frac{m_{e}}{m_{i}}}
\]
Unless the neutral density $n_{n}$ and electron temperature $T_{e}$
in the MC interior are fairly low, the above parameter is not larger
than one, so the density depletion, $\Delta n_{i}/n_{i}\simeq\alpha/t_{0}\ll1$,
remains insignificant during the MC travel through the shock precursor.
This estimate validates our assumption $n_{CR}\ll n_{i}$ and thus
the solution given by eq.(\ref{eq:psiOftauandalpha}).

The weak effect of accelerated CR protons on the MC ion density does
not mean that the charge neutralizing electric field also remains
weak. The electric field can be determined from eqs.(\ref{eq:Eofxt},\ref{eq:Voft},\ref{eq:psiOftauandalpha}),
so we can write it as follows

\[
E\left(x,t\right)\simeq\frac{m_{i}}{e}a\nu_{in}^{2}\frac{x\alpha}{\left(t_{0}-t\right)^{2}}\left[1+\frac{\alpha}{t_{0}-t}\right]
\]
where $E$ is given in physical units, while $x$, $t_{0}$ and $t$
are still dimensionless, stemming from the expansion of $\psi(\alpha,t)$
for small $\alpha/t_{0}$. The second term in the brackets corresponds
to the ion contribution to electric field generation (second term
in the brackets in eq.{[}\ref{eq:Eofxt}{]}). As long as $\Delta n_{i}/n_{i}\simeq\alpha/t_{0}\ll1$,
it can be neglected. The electric field reaches its maximum at the
edge of MC at the moment of subshock encounter, $t=0$. It can be
represented as 

\begin{equation}
E_{{\rm max}}\simeq\frac{m_{e}}{e}u_{1}\nu_{ei}\frac{n_{CR}^{0}}{n_{i}}\label{eq:Emax}
\end{equation}
There are two potentially significant effects of the electric field.
First, it may cause a runaway acceleration of thermal electrons. To
assess this possibility, one needs to compare $E_{{\rm max}}$ with
a critical (Dreicer) field, above which many electrons from the thermal
Maxwellian gain more energy from the field between collisions than
they loose after the collisions \citep{pitaevskii1981physical}. From
the last formula we obtain

\begin{equation}
\frac{E_{{\rm max}}}{E_{{\rm crit}}}\sim\frac{u_{1}}{V_{Te}}\frac{n_{CR}^{0}}{n_{e}}\label{eq:Emax2crit}
\end{equation}
We observe that the collision frequency quite naturally cancels out
in this ratio. It thus depends only on the ratio of the convective
CR flux to that of the thermal electrons. It should also be noted
that to produce a significant effect, $E_{{\rm max}}$ does not need
to be close to $E_{{\rm crit}}$. Even if the above ratio is low,
an exponentially small number of runaway electrons \citep{Gurevich61RunAway}
on the tail of their velocity distribution can initiate an ionization
process. Once started, it may develop into a gas breakdown at fields
significantly below the impact ionization threshold \citep{GurevichPairs00}.
The runaway breakdown, actively studied in terrestrial thunderstorms,
requires a seed population of fast electrons, sporadically produced
by ionizing CRs \citep{GurevichUFN2001}. In an MC ahead of an SNR
shock such population is readily available, e.g., from shock accelerated
electrons and secondary electrons inside the MC. Another possible
effect associated with the runaway process is an electromagnetic cascade
that may result in a pair production which would further increase
the number of positrons injected into the DSA. These phenomena, expected
to occur in strong electric fields, will react back on the field generation
by increasing the neutralizing current. The runaway electrons, in
particular, may carry most of that current. 

The electrostatic potential that obviously has a maximum in the middle
of the MC may partially screen the MC interior from the penetrating
CR, particularly a low-energy and, therefore, more intense part of
their spectrum. The maximum potential is proportional to the MC half-length,
$a.$ To determine the CR penetration into the MC, one needs to compare
this potential with the proton rest energy, $m_{p}c^{2}$. So, from
eq.(\ref{eq:Eofxt}) we obtain

\begin{equation}
\frac{e\phi_{{\rm max}}}{m_{p}c^{2}}\sim\frac{a}{1pc}\frac{u_{1}}{c}\frac{n_{CR}}{1cm^{-3}}\left(\frac{1eV}{T_{e}}\right)^{3/2}\label{eq:fiMAX}
\end{equation}
Similarly to the comparison of the electric field with the critical
Dreicer field in eq.(\ref{eq:Emax2crit}), the last estimate also
places the field potential in a critical range, this time, possibly
close to the relativistic proton energy. A one-parsec size MC does
not appear implausible, as it would be of the order of the gyroradius
of a PeV (knee energy) proton. Such an MC occupies only a $u_{1}/c\ll1$-
fraction of the entire CR precursor, assuming Bohm diffusion regime.
The above estimates indicate that the electric field may grow strong
enough to react back on the penetration of low-energy CRs into the
MC and neutralization of the CR charge by a plasma return current.
These aspects of the CR-MC interaction require a separate study. Here,
we assume the relevant parameters constrained as to keep the ratios
in eqs.(\ref{eq:Emax2crit}) and (\ref{eq:fiMAX}) significantly smaller
than one. 

\section{Spectrum of shock-accelerated positrons\label{sec:Spectrum-of-shock-accelerated}}

In Sec.\ref{subsec:Electrodynamics-of-CR-MC} we have considered relevant
MC processes driven by penetrating CRs. We have found the CR density,
MC electrostatic potential, and ion outflow velocity increasing explosively,
$\propto\left(t_{0}-t\right)^{-1}$, towards the subshock encounter.
Therefore, the positron expulsion from the MC will culminate at the
time of encounter, thus peaking their injection into the DSA process
discussed further in this section. 

\subsection{Positron Injection into DSA\label{subsec:PositronInjection}}

Being interested in a particle injection from many MCs, occasionally
crossing the shock, we may consider the expelled positrons as \emph{injected}
into the DSA at time-averaged rate $Q\left(p,x_{MC}\right)$. It decays
sharply with $x_{MC}$, the distance from the subshock, according
to eq.(\ref{eq:NcrOfx}) which is more convenient to use here than
its time-dependent analog, given in eq.(\ref{eq:NcrOft}). In what
follows, we will write $x$, instead of $x_{MC}$ which should not
cause any confusion with the notation of Sec.\ref{sec:InteractionOfCRwithMC}. 

A momentum distribution of injected positrons is determined by the
history of their production in, and outflow from, an MC. At large
distances from the subshock, only the most energetic CRs penetrate
the MC, while low-energy CRs do not reach it. On the other hand, short
before shock crossing, the low-energy CRs cannot freely penetrate
the MC, because of the induced electric field. Thus, bearing in mind
that positrons receive only a few percent of the energy of parent
protons, it is not unreasonable to expect $Q\left(p\right)$ having
a relatively broad maximum near or somewhat below the momentum $e\phi_{{\rm max}}$/c,
eq.(\ref{eq:fiMAX}). Given relatively cold, e.g., $T_{e}\sim100$K
electrons in the MC, this maximum is likely to be in a sub-GeV range.
Here we orient ourselves towards a cold neutral medium with $n_{{\rm H}}\gtrsim30{\rm cm^{-3}}$
and the filling factor $f_{V}\sim0.01$ \citep{Chev03,Draine2010BookISM}.
The value of $n_{{\rm H}}f_{V}$ may substantially exceed its counterpart
in the ambient plasma. 

Positrons, generated in CR-MC gas collisions are confined in the MC
for a time $\tau_{{\rm conf}}\sim a^{2}/\kappa\left(p\right),$ to
be compared with the precursor crossing time (also CR acceleration
time) $\tau_{a}\sim\kappa\left(p_{{\rm {\rm max}}}\right)/u_{1}^{2}\sim L_{CR}/u_{1}$.
Here $p$ and $p_{{\rm max}}$ denote the positron and maximum CR
(proton) momentum, respectively. Strictly speaking, the particle diffusivities
$\kappa$ are different, as they refer to different media (MC and
ionized CR precursor). Nonetheless, for a simple estimate below, we
may adopt the Bohm scaling for both. If $\tau_{{\rm conf}}/\tau_{a}\gg1$,
which translates then into $p/p_{{\rm max}}\ll a^{2}/L_{CR}^{2}$,
low-energy secondary positrons accumulated in the MC over the precursor
crossing time, will stay inside the MC. Therefore, they will avoid
the DSA process, along with the most of negatively charged particles.
Indeed, a strongly weakened subshock engulfs the MC without shocking
its material over the MC crossing time \citep{Draine2010BookISM,Inoue12}.
To a certain extent, this also relates to the positively charged secondaries
and spallation products, diffusively trapped in the MC. Therefore,
they do not develop spectra similar to that of the positrons. 

Two considerations help to elaborate the above constraint on the positron
momentum $p$. First, a significant modification of the shock structure
requires a proton cut-off momentum $p_{{\rm max}}\gtrsim1$ TeV, while
our interest in positrons is limited to $p<500$ GeV (data availability).
Second, the positrons receive, on average, only about 3 percent of
the energy of parent CRs. These considerations render $p/p_{{\rm max}}\ll1$
as a strong inequality. We, therefore, conclude that except for very
small MCs, the condition $p/p_{{\rm max}}\ll a^{2}/L_{CR}^{2}$ is
fulfilled and most of the early generation of positrons, produced
by high-energy protons, will stay inside the MC. However, in parameter
regimes when the electric field is very strong, this conclusion may
be violated but we assume that it is not. 

On entering the subshock~proximity, the CR number density sharply
increases by GeV particles. To some extent, these particles are screened
by an MC electric field which reaches its maximum at the MC's edge.
Therefore, they generate secondary $e^{\pm}$ and, for that matter,
$\bar{p}$, at the periphery of the MC. The edge electric field then
expels positively charged secondaries ($e^{+}$) and sucks in negatively
charged ones, such as $e^{-}$ and, to some extent, $\bar{p}$, even
though they are considerably more energetic for kinematic reasons.
Based on the calculation of the field in Sec.\ref{subsec:Electrodynamics-of-CR-MC},
the typical energy of expelled positrons should not exceed $1-2$
GeV. This estimate is consistent with that presented earlier in this
subsection. 

Now we turn to the acceleration of injected positrons. It should be
noted, however, that some secondary negatively charged particles,
such as $\bar{p},$ can still be injected along, particularly if the
MC is sufficiently small and the maximum field potential is thus not
high enough to suck them in. In this paper, however, we do not consider
their contribution to the integrated CR spectrum produced in an SNR
of the type considered. Such consideration would require us to address
the question of the MC distribution in size. 

\subsubsection{Shock Acceleration of Positrons\label{subsec:Diffusive-Acceleration-of}}

Upon expulsion from an MC, positrons undergo the DSA. As the shock
is strongly modified, the acceleration starts in its precursor. Because
of the flow convergence, $\partial u/\partial x\neq0$, particles
gain energy without even crossing the subshock \citep{MD06}. On the
other hand, most of the positrons are released from the MC near the
subshock. Thus, at lower energies, their spectrum will be dominated
by the subshock compression ratio, $r_{s}=u_{0}/u_{2}$ rather than
by the precursor precompression, $u_{1}/u_{0}$. Therefore, the spectral
index must be $q=q_{s}\equiv3r_{s}/\left(r_{s}-1\right)$ and the
spectrum $f_{e^{+}}\propto p^{-q_{s}}$. 

By gaining energy, particles sample progressively larger portions
of shock precursor with higher compression ratios, Fig.\ref{fig:Flow-profile-near},
which makes their spectrum harder. On the other hand, as they also
need to diffuse across larger portions of the precursor, the acceleration
slows down which makes the spectrum softer. Asymptotically, these
trends balance each other, but the balance is critically supported
by accelerated protons, and their pressure needs to be included in
the equations for the shock structure. In the case of very strong
shocks ($M\to\infty$) with sufficiently high maximum energy, a universal
spectrum $p^{-3-\sigma/2},$ establishes \citep{MDru01}. Here $\sigma$
is the index of particle diffusivity, $\kappa\propto p^{\sigma}$
with $\sigma=1$ for Bohm diffusion. 

Consider now the ratio of positron spectrum to the spectrum of electrons
produced in unmodified strong shocks with a typical spectrum $\propto p^{-4}$.
This ratio, that is $p^{4}f_{e^{+}}\left(p\right)$, will have a decreasing
branch at low momenta, since $f_{e^{+}}\left(p\right)\propto p^{-q_{s}}$
with $q_{s}>4$, and an increasing branch at high momenta, where the
positron spectral index tends to 3.5. The remainder of this section
deals with the calculation of positron spectrum described above, by
solving the diffusion-convection equation, and comparisons with the
AMS-02 data. 

It is convenient to place the upstream medium in the $x>0$ half-space,
with the subshock at $x=0$, but use positive quantities in describing
the flow velocity. In the subshock frame, the physical flow velocity
starts from $-u_{1}$ at $x=\infty$, decreases gradually to its value
$-u_{0}$ just ahead of the subshock, and then jumps to its downstream
value $-u_{2}$: $0<u_{2}<u_{0}\leq u\left(x\right)<u_{1}$, Fig.\ref{fig:Flow-profile-near}.
We will ignore the inclination of the magnetic field line to the shock
normal (i.e. $x-$ direction), which can be effectively included by
redefining the diffusion coefficient \citep{Drury83}.

Although the dynamics of an individual MC is essentially time dependent
(Sec.\ref{subsec:Electrodynamics-of-CR-MC}), we are interested in
an average positron input from an ensemble of MCs. Therefore, we consider
a steady state problem with a stationary injection of positrons at
the subshock. The injection rate is then given by a time averaged
source $Q\left(x,p\right)$, discussed in Sec.\ref{subsec:PositronInjection}.
The distribution of accelerated positrons is governed by the familiar
convection-diffusion equation

\begin{equation}
u\frac{\partial f}{\partial x}+\kappa\left(p\right)\frac{\partial^{2}f}{\partial x^{2}}=\frac{1}{3}\frac{\partial u}{\partial x}p\frac{\partial f}{\partial p}-Q\left(p,x\right),\label{eq:DC1}
\end{equation}
with a standard normalization of the positron number density: 

\[
n_{e^{+}}\left(x\right)=\int f\left(x,p\right)p^{2}dp,
\]
The momentum dependence of the positron diffusion coefficient $\kappa$
can be taken to be in an ultrarelativistic Bohm regime, $\kappa=\kappa^{\prime}p,$
with $\kappa^{\prime}=const.$

First, we consider the solution to eq.(\ref{eq:DC1}) for moderate
values of $p$, assuming that the positron diffusion length, $\kappa\left(p\right)/u_{0}\ll L_{CR}$,
where, $L_{CR}\sim\kappa\left(p_{{\rm max}}\right)/u_{1}$ is the
precursor scale, determined by the maximum energy of \emph{accelerated
protons}. Hence, we may expand the flow velocity upstream, $u\left(x\right)=u_{0}+u^{\prime}x$,
for $x\ge0$ with $u^{\prime}=const$. At the same time, we will focus
on particle momenta that are higher than the injection momentum, so
we drop the injection term $Q$ in eq.(\ref{eq:DC1}) and include
its effect on the solution in form of normalization of $f$. In particular,
the value $f\left(x=0,p=p_{{\rm inj}}\right)$, where $p_{{\rm inj}}$
is defined as $Q\left(p>p_{{\rm inj}}\right)=0,$ can be expressed
through injection rate $Q$ approximately as 

\begin{equation}
f\left(0,p_{{\rm inj}}\right)\approx\frac{1}{u_{0}}\int_{0}^{\infty}Qdx,\label{eq:fatpinj}
\end{equation}
We have assumed here that, in a steady state considered, injection
is balanced by convection at low momenta. For that reason, we have
neglected diffusion and acceleration terms, according to $\kappa/u_{0}l_{inj}\ll1$
and $u_{1}l_{inj}/L_{CR}u_{0}\ll1$. More accurate determination of
the normalization is not worth the effort, as there are larger uncertainties
in the value of $Q$, associated with our limited knowledge of the
MC density, for example. The main objective here is to determine the
spectral shape of $e^{+}$ which does not depend on the normalization,
as the shock modification is produced by protons, not positrons. 

To lighten notation, we make use of invariant properties of Eq.(\ref{eq:DC1}),
and replace $\kappa\left(p\right)=\kappa^{\prime}p\to p$, which can
easily be reversed by the transform $p\to\kappa$, when the equation
is solved \footnote{For pedantic readers, we use here units for $p$ in which $\kappa^{\prime}=1$.}.
Note that a more general scaling of $\kappa$ with $p$, such as $\kappa\propto p^{\sigma}$
can also be accommodated by a simple change of variables/coefficients.
Adhering to the Bohm scaling, and taking all the above considerations
into account we rewrite eq.(\ref{eq:DC1}) as

\begin{equation}
\left(u_{0}+u^{\prime}x\right)\frac{\partial f}{\partial x}+p\frac{\partial^{2}f}{\partial x^{2}}=\frac{1}{3}u^{\prime}p\frac{\partial f}{\partial p}\label{eq:DC2}
\end{equation}

This equation can be readily solved by applying a Laplace transform

\[
f_{\lambda}\left(p\right)=\int_{0}^{\infty}f\left(p,x\right)e^{-\lambda x}dx,
\]
which yields

\begin{equation}
u^{\prime}\left(\lambda\frac{\partial f_{\lambda}}{\partial\lambda}+\frac{p}{3}\frac{\partial f_{\lambda}}{\partial p}\right)-\left(u_{0}\lambda+\lambda^{2}p-u^{\prime}\right)f_{\lambda}=-\left(u_{0}+\lambda p\right)f_{0}-pf_{0}^{\prime}\label{eq:LaplTrEq}
\end{equation}
Here we denoted $f_{0}\left(p\right)=f\left(x=0,p\right)$ and $f_{0x}\left(p\right)=\left.\partial f/\partial x\right|_{x=0+}$.
The last two functions are related through a jump condition at the
subshock at $x=0$. Integrating eq.(\ref{eq:DC1}) across the subshock,
and taking into account the downstream stationary solution $f\left(p,x\right)=f_{0}\left(p\right)$,
$x<0$, we obtain

\begin{equation}
f_{0x}=\frac{\Delta u}{3}\frac{\partial f_{0}}{\partial p}\label{eq:JumCond}
\end{equation}
The solution of eq.(\ref{eq:LaplTrEq}) can be found, by integrating
along its characteristics on the $\lambda,p$ plane, $\lambda/p^{3}=const$,
and using the jump condition in eq.(\ref{eq:JumCond})

\begin{equation}
f_{\lambda}\left(p\right)=\frac{3}{u^{\prime}p^{3}}e^{\psi_{\lambda}\left(p,p\right)}\int_{p}^{\infty}p^{\prime2}dp^{\prime}e^{-\psi_{\lambda}\left(p^{\prime},p\right)}\left[\left(u_{0}+\lambda\frac{p^{\prime4}}{p^{3}}\right)f_{0}\left(p^{\prime}\right)+\frac{\Delta u}{3}p^{\prime}\frac{\partial f_{0}\left(p^{\prime}\right)}{\partial p^{\prime}}\right]\label{eq:LapTrSol}
\end{equation}
where

\begin{equation}
\psi_{\lambda}\left(p^{\prime},p\right)=\frac{\lambda}{u^{\prime}}\frac{p^{\prime3}}{p^{3}}\left(u_{0}+\frac{3}{7}\lambda\frac{p^{\prime4}}{p^{3}}\right).\label{eq:PsiLamda}
\end{equation}
Using this solution, the function $f\left(x,p\right)$ can be found
by inverting the Laplace transform:

\begin{equation}
f\left(x,p\right)=\frac{1}{2\pi i}\int\limits _{-i\infty+b}^{i\infty+b}e^{\lambda x}f_{\lambda}\left(p\right)d\lambda\label{eq:InvLapl}
\end{equation}
where the constant $b$ must be taken larger than the real parts of
all the singularities, $\lambda_{s}$, of $f_{\lambda}$ on the $\lambda$
plane, $b>\Re\lambda_{s}$. Clearly, a formal solution of eq.(\ref{eq:DC2})
given by eqs.(\ref{eq:LapTrSol}) and (\ref{eq:InvLapl}) still depends
on an unknown function $f_{0}\left(p\right)$, the spectrum at the
subshock. This is because out of the two boundary conditions required
to solve eq.(\ref{eq:DC2}), we used only one, given by eq.(\ref{eq:JumCond})
which connects $f_{0x}$ with $f_{0}\left(p\right)$. The second condition
is $f\left(x,p\right)\to0$ for $x\to\infty$. To fulfil it, all possible
singularities of $f_{\lambda}$ should be limited to the half-plane
$\Re\lambda<0$. 

An inspection of the integrand in eq.(\ref{eq:LapTrSol}) shows that,
under a proper behaviour of $f_{0}\left(p\right)$ at $p\to\infty$,
$f_{\lambda}$ is bounded for $\lambda>0$. So, we focus on a pole
at $\lambda=0$ and upon extracting the term 

\[
f_{\lambda}\sim\frac{S\left(p\right)}{\lambda}
\]
from eq.(\ref{eq:LapTrSol}), the condition $S\left(p\right)=0$ will
need to be imposed. To calculate $S$$\left(p\right)$, let us expand
$f_{\lambda}$ in a series of $u^{\prime}$. Physically, $u^{\prime}\sim u_{1}^{2}/\kappa\left(p_{{\rm max}}\right)$,
where $p_{{\rm max}}$ is the maximum energy of CR protons, shaping
the flow profile upstream by their pressure. The positron momenta,
we are considering here, are much lower, so we may take a limit $p_{{\rm max}}\to\infty$,
that is $u^{\prime}\to0$. A more specific constraint on $u^{\prime}$
will emerge below. Observe that the behavior of $f_{\lambda}$ at
$\lambda\to0$ is controlled by the contribution of large $p^{\prime}$
in the phase function $\psi_{\lambda}$ that can be obtained by expanding
$S$ in small $u^{{\rm \prime}}.$ The first two terms of this expansion
yield the following equation for $f_{0}\left(p\right)$ 

\begin{equation}
S\left(p\right)\approx f_{0}+\frac{p}{q_{s}}\frac{\partial f_{0}}{\partial p}-\frac{u^{\prime}p^{2}}{3u_{0}^{2}}\left(1+\frac{6}{q_{s}}+2\frac{p}{q_{s}}\frac{\partial}{\partial p}\right)\frac{\partial f_{0}}{\partial p}=0\label{eq:SofP}
\end{equation}
Here $q_{s}=3u_{0}/\Delta u$ is a spectral index corresponding to
the subshock compression, $q_{s}=3r_{s}/\left(r_{s}-1\right)$, where
$r_{s}=u_{0}/u_{2}$. As expected, for $u^{\prime}\to0$ we obtain
from eq.(\ref{eq:SofP}) a familiar test-particle solution, $f_{0}\propto p^{-q_{s}}$.
It holds up for the finite $u^{\prime}$ but for relatively low momenta,
$u^{\prime}\kappa\left(p\right)/u_{0}^{2}\ll1$, or $p/p_{{\rm max}}\ll u_{0}^{2}/u_{1}^{2}$.
We have returned to the physical units by replacing $p\to\kappa\left(p\right)$
and used the above estimate for $u^{\prime}$. In fact, the function
$S\left(p\right)$ is expanded in $\alpha p<1,$ where 

\begin{equation}
\alpha=\frac{u^{\prime}}{3u_{0}^{2}}\label{eq:ApphaPar}
\end{equation}
(not to be confused with $\alpha$ in Sec.\ref{subsec:Electrodynamics-of-CR-MC}).
So, the first two terms in eq.(\ref{eq:SofP}) represent the limit
$\alpha p\to0$, while the other terms yield the first order correction
in this variable. We will use this correction to match the solution
of eq.(\ref{eq:SofP}) for $f_{0}\left(p\right)$ with an exact solution
of eq.(\ref{eq:DC2}) in the region of large $p$, where it becomes
independent of the subshock compression. Note that the latter is determined
by the scale and flow precompression in the CR shock precursor. 

To solve eq.(\ref{eq:SofP}) we introduce a new independent variable 

\[
y=1/4\alpha p
\]
and rewrite the equation as follows

\begin{equation}
\frac{d^{2}f_{0}}{dy^{2}}+2\left(1-\frac{q_{s}+2}{4y}\right)\frac{df_{0}}{dy}-\frac{2q_{s}}{y}f_{0}=0\label{eq:f0Eq}
\end{equation}
It is convenient to transform the last equation to a canonical (in
this case Whittaker) form by introducing a new dependent variable
instead of $f_{0}\left(y\right)$

\[
g=f_{0}e^{y-\left(q_{s}+2\right)\ln y/4}
\]
which obeys the equation

\begin{equation}
\frac{d^{2}g}{dy^{2}}-Qg=0\label{eq:WKBeq}
\end{equation}
where 

\[
Q=1+\frac{1}{2y}\left(3q_{s}-2\right)+\frac{q_{s}+2}{16y^{2}}\left(q_{s}+6\right)
\]
Since we will use the $y\gg1$ asymptotic limit of this equation,
instead of expressing the solution of eq.(\ref{eq:WKBeq}) through
Whittaker functions, we apply the WKB approximation. For the same
reason, the $1/y^{2}$- term in the last expression can be omitted.
Moreover, for $y>0$ eq.(\ref{eq:WKBeq}) has no turning points ($Q\neq0,$
since $q_{s}>4$), the following solution can be used for all $y\gtrsim1$,
and it tends to an exact one for $y\gg1$

\[
g=C_{1}Q^{-1/4}\left[e^{\int\sqrt{Q}dy}+D_{1}e^{-\int\sqrt{Q}dy}\right],
\]
where $C_{1}$ and $D_{1}$ are arbitrary constants. It should be
noted that as $Q\sim1$ for the values of $y\gtrsim1$, where we will
match this solution to the high momentum solution that we obtain below,
both linearly independent solutions in the last formula are still
of the same order. This situation is different from more customary
WKB analyses where $Q\gg1$ and the matching procedure consists in
linking linearly independent solutions of the same equation (\ref{eq:WKBeq}).
One of them becomes subdominant and cannot be matched without continuing
to the complex $y$- plane (so-called Stokes phenomenon). By contrast,
we match here solutions of different equations, that is eqs.(\ref{eq:DC2})
and (\ref{eq:WKBeq}).

Returning to the original variables $f_{0}$ and $p$, from the last
relation we obtain

\begin{equation}
f_{0}\left(p\right)\approx C_{2}p^{-q_{s}}\left(1+D_{2}e^{-1/2\alpha p}p^{3q_{s}/2-1}\right)\label{eq:f0OfpSubsh}
\end{equation}
where $C_{2}$ and $D_{2}$ are still arbitrary constants. The underlying
physics behind the last result is obvious. For low particle momenta,
corresponding to a small diffusion length $L_{{\rm dif}}\sim\kappa/u_{0}$,

\begin{equation}
\alpha p=\frac{u^{\prime}\kappa\left(p\right)}{3u_{0}^{2}}\sim\frac{L_{{\rm dif}}\left(p\right)}{L_{CR}}\frac{u_{1}}{u_{0}}\ll1,\label{eq:AlphaTimesP}
\end{equation}
particles 'feel' only the subshock compression, so their spectral
index is close to $q_{s}=3u_{0}/\left(u_{0}-u_{2}\right)$ (first
term in eq.{[}\ref{eq:f0OfpSubsh}{]}). With growing momentum, $L_{{\rm dif}}$
also grows and particles sample progressively larger portions of the
shock precursor, thus feeling higher flow compression. Their spectrum
becomes harder, which is reflected in the second term in eq.(\ref{eq:f0OfpSubsh})
that begins to dominate at larger $p$. However, by the nature of
the expansion in $\alpha p<1$, this solution cannot be continued
to momenta $\alpha p\gtrsim1$, and has to be matched to a proper
solution of eq.(\ref{eq:DC2}). 

An exact asymptotic solution valid in the regime of strong shock modification,
$u_{1}\gg u_{0}$, and for large $p$ is readily available for the
\emph{proton} spectrum, \citep{m97a}. Its \emph{positron} counterpart
must follow the proton spectrum at ultrarelativistic rigidities and
can be extracted from the asymptotic solution. However, it appears
easier and more persuasive to obtain the positron spectrum directly
from the general convection-diffusion eq.(\ref{eq:DC1}), using a
resolving substitution used in the asymptotic solution. We only need
to specify the flow profile $u\left(x\right)$. In the above references,
$u\left(x\right)$ was self-consistently obtained from the momentum
flux conservation across the shock precursor. Here, the positron solution
is essentially a test-particle one which, however, must have the same
asymptotics as the proton solution for $p\gg m_{p}c$. For the proton
spectrum, the linear $u\left(x\right)$ approximation, that also reduces
eq.(\ref{eq:DC1}) to (\ref{eq:DC2}), is acceptable for $p\ll p_{{\rm max}}$
in the Bohm regime, $\kappa\propto p$ \citep{MDru01}, which we adopt
here. Physically, these particles fall into an intermediate energy
range and sample larger flow compressions than that of the subshock
but smaller than the total compression. Therefore, the flow profile
can be approximated by a linear function of $x$.

The following substitution resolves eq.(\ref{eq:DC1})

\begin{equation}
f\left(x,p\right)=F\left(p\right)e^{-q_{b}\left(p\right)\Psi\left(x\right)/3\kappa\left(p\right)}\label{eq:DCasSol}
\end{equation}
which can be shown by direct substitution. Here we have introduced
the flow potential $\Psi$, according to $u=\partial\Psi/\partial x,$
which for the linear flow profile can be represented as

\begin{equation}
\Psi=\frac{u_{0}^{2}}{2u^{\prime}}\left(1+\frac{u^{\prime}}{u_{0}}x\right)^{2}\label{eq:FlowPotDef}
\end{equation}
The spectral index $q_{b}\left(p\right)$ is defined in a standard
way: 

\[
q_{b}=-\frac{p}{F}\frac{\partial F}{\partial p}
\]
Note that unlike $f_{0}\left(p\right)$, used before, $F\left(p\right)$
is not exactly the spectrum at the subshock since $\Psi\left(0\right)\neq0$.
It is easy to verify that the solution in eq.(\ref{eq:DCasSol}) satisfies
eq.(\ref{eq:DC2}) for the following choice of $q_{b}$ and $F$: 

\begin{equation}
q_{b}\left(p\right)=\frac{7}{2}\frac{p^{7}}{p^{7}-p_{0}^{7}}\;\;\;{\rm and}\;\;\;F\left(p\right)=A\left(p^{7}/p_{0}^{7}-1\right)^{-1/2}\label{eq:qbandF}
\end{equation}

The arbitrary constant $p_{0}<p$ in this solution may be used for
matching purposes. It is, however, clear that $p_{0}$ should be close
to a matching momentum, in which vicinity the solutions given by eqs.(\ref{eq:f0OfpSubsh})
and (\ref{eq:DCasSol},\ref{eq:qbandF}) coincide. Indeed, as $f_{0}\left(p\right)$
very quickly becomes scale-invariant, that is $f_{0}\propto p^{-3.5}$
for $p>p_{0}$, $p_{0}$ should be well inside the overlapping region
between the two asymptotics, just to make a smooth matching possible.
For the comparison with the AMS-02 data below, it is important to
realize that $p_{0}$ also depends on the degree of flow modification
through the parameter $\alpha$ in eq.(\ref{eq:ApphaPar}) that, in
turn, enters the low-momentum solution in eq.(\ref{eq:f0OfpSubsh}).
The normalization constant $A$ in eq.(\ref{eq:qbandF}) remains arbitrary
at this point, which we will also use for matching. Altogether, we
have three free parameters to adjust for matching: $B$ in eq.(\ref{eq:f0OfpSubsh}),
and $A$ with $p_{0}$ in eqs.(\ref{eq:DCasSol},\ref{eq:qbandF}).
To minimize the number of matching parameters, we (temporarily) scale
$p_{0}$ out of the problem:

\[
s=\frac{p}{p_{0}}
\]
Using this variable, and combining eqs.(\ref{eq:f0OfpSubsh}), (\ref{eq:DCasSol})
and (\ref{eq:qbandF}), we obtain the following compound solution

\begin{equation}
f_{0}\left(s\right)=\begin{cases}
s^{-q_{s}}+Be^{-1/2\beta s}s^{q_{s}/2-1}, & s\lesssim1\\
A\left(s^{7}-1\right)^{-1/2}e^{-7s^{6}/36\beta\left(s^{7}-1\right)}, & s\gtrsim1
\end{cases}\label{eq:CompoundAsym}
\end{equation}
In place of the parameter $\alpha$ in eq.(\ref{eq:f0OfpSubsh}) we
introduced here a new parameter $\beta=\alpha p_{0}$. We normalized
the low-momentum asymptotics arbitrarily, $\approx s^{-q_{s}}$, bearing
in mind that the actual normalization factor is proportional to the
injection source $Q$. Its intensity, in turn, depends on the MC density,
which remains a free parameter.

\begin{figure}
\includegraphics[bb=0bp 220bp 612bp 750bp,clip,scale=0.6]{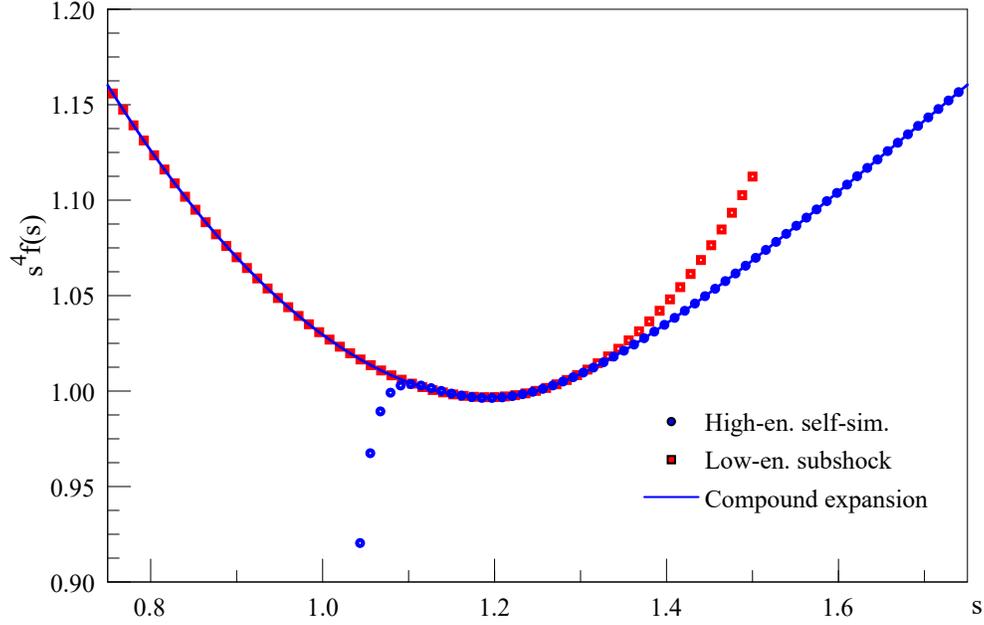}

\caption{Matching of low (dashed line) and high (solid line) momentum solutions
of eq.(\ref{eq:DC2}) given by eq.(\ref{eq:CompoundAsym}). An overlap
region at $s\gtrsim1$ ensures a smooth transition between the two
asymptotics. They deviate from the actual solution at high/low momenta.
The matched asymptotic (compound), uniformly valid solution is shown
by the solid line. The matching parameters in eq.(\ref{eq:CompoundAsym})
are $\beta=0.95$, $B=0.05$ and $A=0.9785.$ The subshock compression,
$r_{s}=3$, so $q_{s}=4.5.$ \label{fig:Matching-of-low}}
\end{figure}

The easiest way of matching the above two expressions is to plot them
and adjust the parameters $A,\;B$ and $\beta$ to make the transition
as smooth as possible. Note that, as we adjust three parameters, the
result is smooth to the second derivative. This means that we match
the normalization, the index and the curvature of the spectrum. The
result is illustrated in Fig.\ref{fig:Matching-of-low}. As the parameter
$p_{0}$ was scaled out of the matching process, the obtained matching
parameters $A,$ $B$ and $\beta$ are valid for a range of $p_{0}$.
By varying $p_{0}$ we will model the time-dependent shock conditions
(degree of its modification, Mach number, and the proton maximum energy).
This flexibility of the compound solution will be useful for the comparison
with AMS-02 data in the next subsection.

To conclude this subsection, a more rigorous matching would address
an intermediate solution expansion asymptotically approaching each
of the solutions given by eq.(\ref{eq:CompoundAsym}). However, the
following argument renders this more elaborate approach unnecessary.
As we mentioned, shock parameters, such as $\alpha$ and $p_{0}$,
slowly change in time, as do the maximum momentum of accelerated protons
and shock Mach number. The positron spectrum should then be obtained
by integrating over the active live time of a source (SNR). This variation
would affect the overall spectrum more significantly than any further
improvement of the matching procedure could.

\subsection{Comparison of AMS-02 data with the solution of convection-diffusion
equation}

The positron energy spectrum, recently published in the form of $e^{+}/\left(e^{+}+e^{-}\right)$
fraction by the AMS-02 team \citep{AMS02_2014} is highly revealing
of the underlying acceleration mechanism. This fraction is almost
certainly invariant under transformations of the individual $e^{\pm}$
spectra due to otherwise very uncertain propagation effects which
often cause disagreement between models. The unique opportunity to
study the acceleration mechanism is in that the AMS-02 data are likely
to be probing into the positron fraction directly in the source. The
difference in charge sign is unlikely to be important en route. We
will further discuss the propagation aspects in the next section. 

Let us break down the leptonic components comprising the positron
fraction into the following three groups: (1,2) positrons and electrons
produced in the SNR under consideration, with $f_{e^{\pm}}$ being
their momentum distributions, and (3) electrons produced in all other
SNRs, with $f_{e^{-}}^{B}$ being a (background) spectrum thereof.
The AMS-02 positron fraction can then be represented by the following
ratio

\begin{equation}
\mathcal{F}_{e^{+}}\equiv\frac{f_{e^{+}}}{f_{e^{-}}^{B}+f_{e^{-}}+f_{e^{+}}}\label{eq:PosFraction}
\end{equation}
We postulate that the background electrons are diffusively accelerated
in strong but unmodified shocks. For the lack of information about
the distances to the sources that contribute to the above positron
fraction, we also assume that the background electrons propagate the
same average distance as $e^{\pm}$ (1,2), so their equivalent spectrum
(were it produced at the $e^{\pm}$ locale (1,2)), can be taken to
be $f_{e^{-}}^{B}\propto p^{-4}$. Because both $f_{e^{\pm}}$ result
from a single shock acceleration, their momentum profiles above the
injection momenta are identical. But the normalization factors are
clearly different. In this paper, we considered only the positron
injection, so the ratio $f_{e^{+}}/f_{e^{-}}$ is a free parameter
that depends on the density of MCs and electron injection efficiency.
On dividing numerator and denominator of the fraction in eq.(\ref{eq:PosFraction})
by $f_{e^{-}}^{B}\propto p^{-4}$, and introducing a new function
$f_{0}\left(p\right)$ and $e^{\pm}$ normalization by the following
relations $f_{e^{+}}=Cf_{0}\left(p\right)f_{e^{-}}^{B}p^{4},$ $f_{e^{-}}=\left(\zeta-C\right)f_{0}\left(p\right)f_{e^{-}}^{B}p^{4}$
instead of eq.(\ref{eq:PosFraction}), we obtain

\begin{equation}
\mathcal{F}_{e^{+}}=\frac{Cf_{0}p^{4}}{1+\zeta f_{0}p^{4}}\label{eq:PosFrac2}
\end{equation}
Here $C$ is a normalization constant that absorbs input parameters
of the model, such as the MC density and their filling factor in the
SNR environment, distance to the SNR, and intensity of the background
electrons, local to the SNR. We are free to adjust the factor $C$
to fit the positron fraction to the AMS-02 data without compromising
the model. The parameter $\zeta$ quantifies the $e^{\pm}$ \emph{combined
}contribution from the SNR, also relative to the background electron
spectrum. In the present model, the parameter $\zeta$, being related
to $C$, is also indeterminate, partly for the above reasons, but
more importantly, because we do not know the number of injected positrons
relative to the number of injected electrons. This number can in principle
be calculated, but the absence of a reliable electron injection theory
is a serious obstacle for such calculations. There is also an implicit
parameter, $p_{0}$, introduced in high-energy part of the positron
distribution in eq.(\ref{eq:qbandF}). By contrast to $C$ and $\zeta$,
$p_{0}$ is a purely technical parameter here. It can be obtained
from a fully nonlinear acceleration theory \citep{m97a,MDru01}, where
the shock structure is calculated self-consistently with the proton
acceleration. As we mentioned earlier, such calculation would require
the following three acceleration parameters: the Mach number, proton
cut-off momentum, and proton injection rate. 

Here, we are primarily interested in lepton acceleration and treat
them as test particles in a shock structure dynamically supported
by the pressure of accelerated protons. Therefore, the role of the
protons is encapsulated in the parameters $p_{0}$ and $r_{s}$ (or,
equivalently, $q_{s}$), that can be recovered from the above references.
As these parameters still depend on the above three shock characteristics,
which change (albeit slowly) in time, the parameter $p_{0}$ also
changes in time according to the Sedov-Taylor blast wave solution
and particle acceleration rate. The resulting spectrum of the positron
fraction can then be obtained by integrating over $p_{0}$ and other
shock parameters, following the approach suggeested, for example,
in ref.\citep{MDSPamela12} in studying the $p$/He anomaly. In this
paper, however, we take a simpler route. First, we fix the subshock
compression, as $r_{s}$ is almost universally consented to be self-regulated
at a nearly constant level in the range $r_{s}=2.5-3$ during the
efficient phase of acceleration. Namely, had $r_{s}$ dropped below
this range, the proton injection would be suppressed, so the shock
modification diminished, thus driving $r_{s}$ towards the unmodified
value of $r_{s}=4$. Conversely, should $r_{s}$ rise above the range
of 2.5-3, the injection will be increased, thus resulting in a stronger
shock modification and reduced $r_{s}$. A possible additional contribution
from MCs upstream may alter this simple feedback loop. To avoid further
complications, we do not include it but note that the enhanced proton
injection facilitates the efficient (nonlinear) shock acceleration
regime.

The effect of changing $p_{0}$ can be modeled by averaging the calculated
positron fraction over a range of $p_{0}$ variation during the acceleration
history. This range can be inferred from a set of solutions of full
nonlinear acceleration problem presented for different Mach numbers
and $p_{{\rm max}}$, e.g., in Fig. 5 of ref. \citep{MDru01}. As
may be seen from it, the spectral index of strongly modified shocks
crosses $q=4$ (which corresponds to the minimum in $\mathcal{F}_{e^{+}})$
in the range of 5-10 GeV/c, weakly depending on the above acceleration
parameters. Based on our matching procedure in the preceding subsection,
the minimum should be at $\approx1.2\cdot p_{0}$. So, we will compare
our results with the AMS-02 data by averaging $\mathcal{F}_{e^{+}}$
over the range of $p_{0}$, suggested by the nonlinear shock acceleration
theory. However, it is instructive to start with a fixed value of
$p_{0},$ using two different sets of constants $C$ and $\zeta$,
characterizing two different $e^{\pm}$ ratios in the source. 

\begin{figure}
\includegraphics[bb=0bp 220bp 612bp 750bp,clip,scale=0.6]{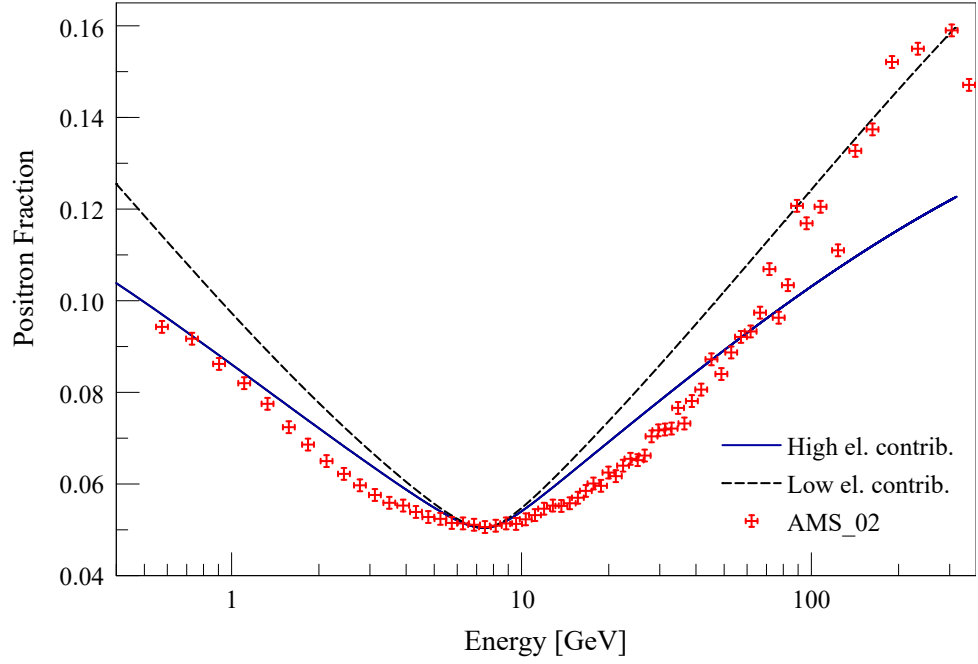}

\caption{Positron fraction, represented as a ratio of $e^{+}$ spectrum to
the sum of $e^{+}$ and $e^{-}$, as given in eq.(\ref{eq:PosFrac2}).
A fit of the AMS-02 data to the solution of eq.(\ref{eq:DC2}) given
by eq.(\ref{eq:CompoundAsym}) with $\beta=0.95$, $p_{0}=6.33$,
$A=0.9785$, $B=0.05$ and the mix of species represented by eq.(\ref{eq:PosFrac2})
is shown for the two sets of normalization and $e^{\pm}$ mixing constants,
$C$ and $\zeta$. They correspond to a high and low electron contribution
to the mix, with $\zeta=9$ and $\zeta=5,$ respectively. To comply
with the AMS-02 at the spectrum minimum, we fixed the normalization
constant at $C=1.45$ and $C=1.25$ for these two cases.\label{fig:PosFrac}}
\end{figure}

Shown in Fig.\ref{fig:PosFrac} is the positron fraction for the matching
parameters $\beta$, $A$ and $B$ indicated in Figs.\ref{fig:Matching-of-low}
and \ref{fig:PosFrac}, for two different combinations of parameters
$C$ and $\zeta$, representing a high ($\zeta=5$) and low ($\zeta=9$)
contribution of the background electrons, respectively. The predicted
saturation level of $\mathcal{F}_{e^{+}}=C/\zeta$ at $p\to\infty$
is 0.16 and 0.25, respectively. The current AMS-02 high energy points
appear to saturate near the lower boundary of this range. The parameter
$p_{0}$ is fixed at $p_{0}=6.33$ in both cases, thus placing the
minimum at $\approx8$GeV/c. This value coincides with the AMS-02
minimum and is well within the range predicted in \citep{MDru01}. 

The first thing to note about Fig.\ref{fig:PosFrac} is that the minimum
is too sharp compared with the AMS-02 data. This is clearly due to
an implausible assumption of a fixed $p_{0}$ that we made to illustrate
the mixing effects of $e^{\pm}$. Since $p_{0}$ varies in time only
within a relatively narrow range, we can model the effect of its variation
by calculating the positron fraction at the minimum and maximum values
of $p_{0}$. The mean value of these two fractions is then a good
proxy for a time integrated spectrum, to be compared with the AMS-02
data. 

The effect of $p_{0}$ variation is shown in Fig.\ref{fig:FinalFit},
which indeed demonstrates a considerably better agreement. A significant
deviation from the AMS-02 data points begins only at high energies,
where they are strongly scattered and, also, have increasingly large
error bars. The theoretical predictions can also be improved in this
area by using the full nonlinear solution, discussed above. Such solution
shows a more gradual transition to the asymptotic $p^{-7/2}$ spectrum
than the one used here. Recall that the latter was based on a linear
profile of the flow velocity upstream. This approximation becomes
inaccurate for high energy particles which reach far upstream, where
the flow velocity saturates at $u=u_{1}$. However improving the full
nonlinear solution, it is unlikely to reconcile with the AMS-02 excess
above the analytical solution in the range $100-300$ GeV.

The present model predicts $\mathcal{F}_{e^{+}}$ to saturate at $\mathcal{F}_{e^{+}}\left(\infty\right)=C/\zeta\approx0.17$,
and $f_{e^{-}}/f_{e^{+}}\approx4.7$ as $p\to\infty$, which is consistent
with the AMS-02 measurements (although errors are significant in this
range). This saturation level is well above the strict upper bound
of 25\% permissible for the SNR contribution to the total postitron
excess. Such limit has been placed in ref.\citep{CholisDMpulsars13},
to avoid conflicts with heavier secondaries accelerated in SNR. This
limitation strictly applies to the acceleration of secondary positrons
generated by $pp$ collisions outside of MCs, where they have no advantages
over heavier secondaries, such as boron, and particularly electrons
and antiprotons. Although it does not restrict the present mechanism
of positron generation, it can be used to constrain the MC density
and filling factor required for the excess. It should be noted that
other studies \citep{Mertsch2009PhRvL,Mertsch2014PhRvD} admit larger
contributions from SNR by using CR diffusivity more rapidly growing
with momentum. The issue is expected to be settled after the AMS-02
data on $\bar{p}/p$ and B/C are published. 

The obtained saturation level is way below 70\%, \emph{predicted}
by the authors of ref.\citep{Cowsik2014ApJ} assuming a discrete distribution
of CR sources (see also \citep{Gaggero2013PhRvL}). Their disagreement
with the results cited above appear to be in part due to the production
of secondaries in the ``cocoon'' region near the SNR, included in
\citep{Cowsik2014ApJ}. However, as was shown in \citep{MetalEsc13},
the near zone of the SNR requires a different approach to CR propagation.
It is based on a CR \emph{self-confinement} supported by the emission
of Alfven waves, rather than commonly used \emph{test particle} propagation.
This is necessary to remain consistent with the well-established idea
of bootstrap acceleration in the SNR shock waves. 

The saturation of the positron fraction in eq.(\ref{eq:PosFrac2})
requires the background electron spectrum $f_{e^{-}}^{B}$ to remain
softer than the $f_{e^{\pm}}$ at high energies. If there were another
electron contribution with a harder than $f_{e^{\pm}}$ spectrum,
but with lower intensity, it would reveal itself at higher energies
and the positron fraction $\mathcal{F}_{e^{\pm}}$ would begin to
decline again, thus creating a maximum at high energies, instead of
leveling off. There are no indications for such additional electron
component as yet, so we do not consider this possibility. Therefore,
the positron fraction in eq.(\ref{eq:PosFrac2}) has no extrema other
than those of the function $f_{0}p^{4}$. As we discussed, apart from
the maximum below the cut-off, it has only one minimum at $\lesssim10$
GeV. We discuss implications of this simple observation below. 

\subsection{Attempts at Interpreting the Data that do not Fit\label{subsec:Interpretations-of-the}}

\begin{figure}
\includegraphics[bb=0bp 220bp 612bp 750bp,clip,scale=0.6]{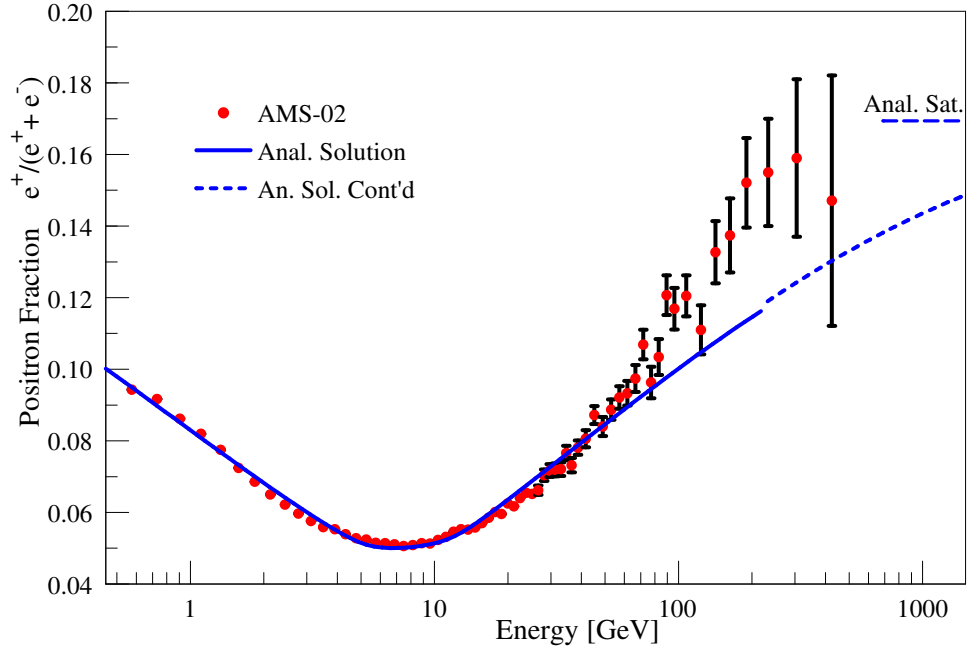}\caption{The same as Fig.\ref{fig:PosFrac} but plotted for an averaged shock
modification: instead of the specified shock modification parameter
$p_{0}=6.33$ used in Fig.\ref{fig:PosFrac}, two different values
of $p_{0}$ are chosen, $p_{0,1}=5.2$ and $p_{0,2}=8.0.$ Shown is
the positron fraction, obtained as an average between those obtained
for $p_{0}=p_{0,1}$ and $p_{0,2}$ (solid line). The dashed line
extends this solution to higher energies, using a simplified calculation
with a fixed value of $p_{0}\approx\left(p_{0,1}+p_{0,2}\right)/2$,
because the effect of $p_{0}$ dispersion is not significant at high
energies. Parameters in eq.(\ref{eq:PosFrac2}) are fixed at $C=0.061$
and $\zeta=0.35$, so that the saturation level predicted by eq.(\ref{eq:PosFraction})
is $\approx0.17$, shown in the upper right corner. AMS-02 error bars
added, where they are significant ($E>30$ GeV).\label{fig:FinalFit} }

\end{figure}

A V-curve representing the analytic solution shown in Fig.\ref{fig:FinalFit}
fits well to the AMS-02 data over slightly more than two decades in
energy. Other than the normalization of $e^{\pm}$, no free parameters,
such as weights of different sources, propagation parameters, etc.,
have been introduced. Only the subshock spectral index, relevant to
the lowest momenta, was determined using a simplified solution of
a full nonlinear acceleration problem, as discussed above. Therefore,
there is a good reason to believe that the V-curve can be continued
using the obtained solution also to higher momenta. Above$\sim70$
GeV the agreement comes to an end, at least for the next $200-300$
GeV. Let us ignore for a moment the growing error bars and take the
data points as they are. Apart from a strong data scattering between
$\sim70-100$ GeV, a distinct rise in the data above the SNR background
(represented by solid and dashed lines) is observed. It may be interpreted
in several ways, both exciting and prosaic ones. We briefly consider
the following three. 

\paragraph{\textbf{Dark Matter or Pulsar Peak}}

Again, ignoring the error bars, taking the decreasing trend between
the two highest energy data points as a reality, and the model essentially
correct, we expect the higher energy points (when available) to return
on the dashed line. Such behavior will make a strong case for the
excess in the 100-300 GeV having nothing to do with the processes
described by the present model. It can then be interpreted as a dark
matter or pulsar contribution with a cutoff at 300-400GeV \citep{Zhang01,DMtheory2009PhRvD,Hooper09,Malyshev2009PhRvD,Profumo2012CEJPh,HooperDM_AMS13,CholisDMpulsars13,Linden2013ApJ,CholisHooper2014PhRvD}.
In this scenarios, the solid and dashed lines would represent an ``astrophysical
background'' to be subtracted from the $e^{\pm}$ spectra to extract
the new signal. Note that this background is quite different from
that normally used for the purpose, e.g. \citep{CholisHooper2014PhRvD,Boudaud2015A&A}.
It is rising rather than falling with the energy, thus allowing for
a more gradual high energy fall-off in the future data, to admit the
dark matter interpretation. More about this scenario can be told when
the error bars shorten, and new data points are available. 

\paragraph*{\textbf{Synchrotron pile-up}}

Webb et al. showed \citep{DruryPileUp84} that, if the lepton spectrum
is harder than $p^{-4}$ below the synchrotron cut-off, particles
accumulate in this energy range, and the spectrum flattens before
it cuts off. As the SNR spectrum, shown in Fig.\ref{fig:FinalFit},
is essentially $p^{-3.5}$, the AMS-02 excess in the 100-300 GeV range
can, in principle, be accounted for by this phenomenon. However, this
energy is too low for the typical ISM magnetic field of a view $\mu$G
to balance acceleration and losses. At the same time, the magnetic
field in MCs is usually significantly higher, even though their filling
factor is not large. Therefore, a diffusive trapping time of $100$
GeV leptons in MC may be long enough to enhance the losses significantly
and facilitate the pileup. The magnetic field can also be amplified
by a nonresonant, proton driven instability \citep{Bell04} outside
the MC and, to some extent also in its interior \citep{Reville07}.
The MC electrostatic potential ($\sim GV)$ can hardly enhance the
electron trapping time compared to that of positrons.

\paragraph*{\textbf{Model deficiency}}

Taking the AMS-02 error bars in Fig.\ref{fig:FinalFit} more seriously,
one may alternatively assume that the positron fraction will continue
increasing with a likely saturation at higher energies. The explanations
suggested in the two preceding paragraphs can then be dismissed, and
the deviation from the present theory prediction at $>100$ GeV should
be attributed to the incomplete description of particle acceleration
in a CR modified shock. If true, then essentially no room is left
for additional sources, such as the dark matter annihilation or decay.
The model will need to be systematically improved, which is straightforward,
as the technique for obtaining a more accurate nonlinear solution
is available. It requires solving an integral equation \citep{m97a}
instead of a PDE equation we solved in Sec.\ref{subsec:Diffusive-Acceleration-of}.
Also, the entire V-curve in Fig.\ref{fig:FinalFit} will need to be
recalculated including a self-consistently determined low-energy spectral
index, without a simple matching procedure we used in this paper.
Although we do not expect such improvement to be significant, it will
be done in future work, if the decreasing trend at highest energies
is not confirmed. Another possible contribution to the positron excess
may come from a runaway avalanche and pair generation inside the MC
\citep{GurevichPairs00,GurevichUFN2001}.

\section{Discussion of Alternatives \label{sec:Discussion}}

Although the interpretations of the positron anomaly often appear
plausible (see, e.g., \citep{Panov2013JPhCS,Cowsik2014ApJ} for a
review), they do not form one cohesive picture. The problem seems
to be that very different models fit the data equally well. Indeed,
if we ignore the energy range beyond 200-300 GeV, where even the AMS-02
data remain statistically poor, what needs to be reproduced are two
power-laws (below and above $8$ GeV) and the crossover region, characterized
by the position of the spectral minimum and its width (spectral curvature).
Altogether, the fit thus requires four parameters. Let $\gamma_{1}>\gamma_{2}$
be the spectral indices of the two often assumed \emph{independent
}positron contributions, and $C_{e^{+}}$- their relative weight.
The positron flux is $\Phi_{e^{+}}\propto p^{-\gamma_{1}}+C_{e^{+}}p^{-\gamma_{2}}$.
Dividing $\Phi_{e^{+}}$ by its sum with an electron background spectrum
$C_{e^{-}}^{B}p^{-\gamma_{e}}$, that provides the fourth parameter
$C_{e^{-}}^{B}$, one obtains the positron fraction

\begin{equation}
F_{e^{+}}=\frac{p^{-\beta_{1}}+C_{e^{+}}p^{\beta_{2}}}{C_{e^{-}}^{B}+p^{-\beta_{1}}+C_{e^{+}}p^{\beta_{2}}},\label{eq:DiscFrac}
\end{equation}
that depends on four parameters. Here $\beta_{1}=\gamma_{1}-\gamma_{e}>0$
and $\beta_{2}=\gamma_{e}-\gamma_{2}>0$. Not the same but essentially
equivalent spectrum fitting recipe was suggested in the original AMS-02
publication \citep{AMS02_2014}. Not surprisingly, physically different
models produce good fits as they effectively need to provide just
a correct combination of four \emph{independent} parameters. The question
is then how many of them are ad hoc?

Returning to our analog of eq.(\ref{eq:DiscFrac}) given in eq.(\ref{eq:PosFrac2}),
we note that $f_{0}\left(p\right)$ is determined by the shock history.
So, only two independent parameters, $C,$ and $\zeta$, remain to
be specified. As we have seen in the previous subsection, these parameters
correspond (up to linear transformation thereof) to the positron weight
relative to electrons from the same SNR and the ISM background. One
of them is a true free parameter corresponding to the unknown MC density,
filling factor, and some nearby SNRs that contribute to the positron
fraction. The number of positrons, extracted from an MC \emph{relative}
to the number of \emph{injected electrons}, is possible to calculate
in principle but challenging. There are two major problems with such
calculations. First, the rate at which electrons are injected from
the ambient plasma, regardless of the MCs and positrons, is a long-standing
problem in plasma astrophysics \citep{Levinson92,GMV95}. Second,
the extraction of positrons from the MC may be associated with the
gas breakdown and positron/electron runaway accompanied by the pair
production \citep{GurevichPairs00,DwyerRev12}. As we stated earlier,
the latter phenomena are not addressed in this paper, which places
certain limits on the parameters of MCs considered, thus producing
further uncertainty in the positron normalization.

Sources of positrons other than the secondaries from $pp$ collisions
have also been suggested. These are the radioactive elements of the
SN ejecta \citep{Zirak2011PhRvD}, pulsars, and dark matter related
scenarios \citep{Hooper09,HooperDM_AMS13,CholisDMpulsars13,Profumo2012CEJPh}.
However, these scenarios seem to have enough ``knobs'' to tweak
their ``four parameters''. Some SNR based approaches, e.g., \citep{ErlykWolf2013APh}
directly use the AMS-02 data and the background radio indices \citep{Mertsch2014PhRvD}
to infer the fitting parameters. It is not clear if these indices
are a good proxy for the parent proton indices responsible for the
positron production. The radio indices are known to be highly variable
\citep{Strong2011}. The position of the spectral minimum also needs
to be taken directly from the AMS-02 data. Therefore, the physics
of the spectrum formation remains unclear, and the conclusion about
the likely absence of the dark matter contribution is not well justified.
By contrast, the present model attributes the spectral minimum to
the familiar nonlinear shock structure supported by mildly relativistic
protons. Understanding the minimum validates the model prediction
of the decreasing and increasing branches around it, but only to the
next spectral feature. Such feature indeed emerges at $\sim100$ GeV,
but it is too early to say what it is. It is crucial whether a trend
in this feature towards the present model predictions in Fig.\ref{fig:FinalFit}
is confirmed by the next AMS-02 data release. If it is, the 100-300
GeV feature may have nothing to do with the positron generation in
SNR. Then, it is available for more interesting interpretations, such
as dark matter or pulsar contributions to the positron excess.

Now we return to the question whether other charge-sign effects, known
in the DSA, may produce the $e^{+}/e^{-}$ anomaly. It has been argued
for quite some time \citep{m98,MDSPamela12} that injection of particles,
at least in quasi-parallel shocks, promotes their diversity through
\foreignlanguage{american}{disfavoring} the most abundant species,
i.e., protons. The segregation mechanism is simple. Protons, being
injected in the largest numbers, but not necessarily most efficiently,
still make a dominant contribution to the growth of unstable Alfven
waves in front of the shock. In collisionless shocks, such waves support
the shock transition by enabling the momentum and energy transfer
between upstream and downstream plasmas, when the binary collisions
are absent. In particular, the unstable waves control the particle
injection by transporting them to those parts of the phase space in
shock vicinity where they can cross and re-cross its front, thus undergoing
the Fermi-I acceleration. As these waves are driven resonantly, that
is in a regime in which the wave-particle interaction is most efficient,
they react back on protons also most strongly, as the wave driving
particles. Furthermore, the waves are almost frozen into the local
fluid so, when crossing the shock interface, they also entrain most
particles and prevent them from escaping (or reflecting off the shock)
upstream, thus significantly reducing their odds for injection. Again,
most efficient is namely the proton entrainment, while, e.g., alpha
particles have somewhat better chances to escape upstream and to get
eventually injected. The wave-particle interaction for He is weaker
because of the mismatched wavelengths generated by the protons since
their mass to charge ratios are different. The difference in the charge
sign also contributes in disfavor of protons, but this time through
the sign of the wave helicity they drive. The mechanism is simply
that the particle orbit, spiraling along with the spiral magnetic
field of the wave, has a preferred escape direction along the mean
field that depends on the charge sign, given the field direction \citep{m98}.
So, $\bar{p}$ for example, would have better chances for injection
than $p$ but, as we argued earlier, most of them are likely to be
locked in MCs, and so entrained with the shock flow. 

The arguments concerning difference in $\bar{p}$ and $p$ injection,
equally apply to $e^{-}$ and $e^{+}$ of similar rigidity. Therefore,
positrons would be disfavored in the injection context by a conventional,
wave-particle interaction-based injection mechanism, were they injected
from a thermal pool. However, in this paper, we focused on positrons
released from MCs upstream with energies much higher than the injection
energy of protons from the thermal pool. Therefore, they are scattered
by much longer waves whose spectrum is turbulent and probably mirror-symmetric,
so the helicity-dependent, coherent wave-particle interactions considered
in \citep{m98} are irrelevant. By the same token, electrons could
be injected more easily, but the main problem for them is to reach
gyroradii comparable to the proton-driven wave-lengths. Only in this
case could they be de-trapped from the downstream turbulence and start
crossing the shock. Returning to the positrons, we conclude that,
although the charge-sign effect in their interaction with the CR-driven
(primarily by suprathermal protons) turbulence cannot be ruled out,
its role in the positron injection is unlikely to be significant. 

\section{Conclusions and Outlook\label{sec:Conclusions}}

The objectives of this paper have been a detailed explanation of the
$e^{+}/e^{-}$ energy spectrum and understanding of the charge-sign
dependent particle injection and shock acceleration. The principal
results of our study are:
\begin{enumerate}
\item assuming that an SNR shock environment contains clumps of weakly ionized
dense molecular gas (MC), we investigated the effects of their illumination
by shock accelerated protons before the shock traverses the MC. The
main effects are the following:
\begin{enumerate}
\item an MC of size $L_{{\rm MC}}$ is charged (positively) by penetrating
protons to$\sim\left(L_{{\rm MC}}/pc\right)\left(V_{sh}/c\right)\left(1eV/T_{e}\right)^{3/2}\left(n_{CR}/cm^{-3}\right)$GV,
eq.(\ref{eq:fiMAX})
\item secondary positrons produced in $pp$ collisions inside the MC are
pre-accelerated by the MC electric potential and expelled from the
MC to become a seed population for the DSA
\item most of the negatively charged secondaries, such as $\bar{p}$, along
with electrons and heavier nuclei, remain locked inside the MC
\end{enumerate}
\item assuming that the shock Mach number, the proton injection rate, and
their cut-off momentum exceeds the threshold of efficient acceleration
regime \citep{MDru01}, we calculated the spectrum of injected positrons
and, concomitantly, electrons
\begin{enumerate}
\item the momentum spectra of accelerated leptons have a concave form, characteristic
for nonlinear shock acceleration, which physically corresponds to
the steepening at low momenta, due to the subshock reduction, and
hardening at high momenta, due to acceleration in the smooth part
of the precursor flow
\item the crossover region between the trends in (a) is also directly related
to the change in the proton transport (from $\kappa\propto p^{2}$
to $\kappa\propto p$) and respective contribution to the CR partial
pressure in a mildly-relativistic regime. The crossover pinpoints
the 8 GeV minimum in the $e^{+}/\left(e^{+}+e^{-}\right)$ fraction
measured by AMS-02 
\item due to the nonlinear subshock reduction, the MC crosses it virtually
unshocked so that secondary $\bar{p}$ and, in part, heavier nuclei
accumulated in its interior largely evade shock acceleration
\end{enumerate}
\end{enumerate}
Some important physical aspects of the proposed mechanism have not
been elaborated. These include, but are not limited to, the following
\begin{enumerate}
\item calculation of energy distribution of runaway positrons preaccelerated
in MC before their injection into the DSA
\item calculation of electron injection for this kind of shock environment
\item evaluation of conditions for the runaway gas breakdown in MC with
associated pair production and calculation of the yield of this process
\item escape of secondary antiprotons, generated in outer regions of MC
or with sufficient energy, to the ambient plasma and their subsequent
diffusive acceleration 
\item integration of the present calculations of positron spectra into the
available fully nonlinear DSA solutions 
\item study of the MC interaction with a supersonic flow in modified shock
precursor, bow shock formation and implications for additional particle
injection
\end{enumerate}
Implementation of items (1), (2) and (5) will be particularly useful
when AMS-02 gathers more statistics in the $>10^{2}$GeV range, so
that the positron fraction saturation level can be more accurately
compared with the prediction of the improved model.

\appendix

\section{Propagation of CRs inside MC\label{sec:AppCRinsideMC}}

The spectrum of shock accelerated CRs in the MC interior may be different
from that on its exterior for many reasons. First, if the MC size
is comparable to the shock precursor, then the shock-accelerated particles,
while penetrating the MC from its near side, may quickly escape through
its far side \citep{MDS_APS12}. They escape if Alfven waves, confining
particles to the shock precursor, develop a gap in their power spectrum,
which is due to ion-neutral collisions \citep{KulsrNeutr69}. Furthermore,
an electric field that builds up in response to the CR penetration
will shield the MC from the low-energy CRs. Finally, the magnetic
field may be considerably stronger in the MC than in the shock precursor,
and magnetic mirroring may become relevant as well. 

The CR propagation inside the MC can be treated using a standard pitch
angle diffusion equation with magnetic focusing and electric field
terms:

\begin{equation}
\frac{\partial f}{\partial t}+v\mu\frac{\partial f}{\partial x}-\frac{v}{2B}\frac{\partial B}{\partial x}\left(1-\mu^{2}\right)\frac{\partial f}{\partial\mu}-e\frac{\partial\phi}{\partial x}\left(\mu\frac{\partial f}{\partial p}+\frac{1-\mu^{2}}{p}\frac{\partial f}{\partial\mu}\right)=\frac{\partial}{\partial\mu}D\left(p,\mu\right)\frac{\partial f}{\partial\mu}\label{eq:KinEq}
\end{equation}
Here $\mu=p_{\parallel}/p$, $v\approx c$ and $p$ denote particle
velocity and momentum, respectively, $p_{\parallel}$ is the momentum
projection on the local field direction. An induced electric field
potential $\phi\left(x\right)$, and magnetic field $B\left(x\right)$,
are allowed to slowly (on the gyroradius scale) vary along the coordinate
$x\parallel B$. The pitch-angle diffusion coefficient, $D$, turns
to zero in the $\mu,p$ regions, where the resonant Alfven waves are
evanescent, as mentioned above. 

Consider first the latter case, i.e., a scatter-free (no resonant
Alfven waves) particle propagation into the MC. Looking for a steady
state solution of the above equation with a zero r.h.s., we find

\begin{equation}
f=f_{in}\left(\mathcal{H},I\right),\label{eq:finMCgen}
\end{equation}
where $\mathcal{H}\left(p,x\right)$ and $I\left(p,\mu,x\right)$
are the particle energy and magnetic moment, respectively: 

\begin{eqnarray}
\mathcal{H} & = & c\sqrt{p^{2}+m_{p}^{2}c^{2}}+e\phi\label{eq:Hamiltonian}\\
I & = & \frac{p^{2}}{B}\left(1-\mu^{2}\right).\nonumber 
\end{eqnarray}
Here $f_{in}$ is an arbitrary function of its arguments that must
be determined from the boundary condition at the edge of the MC. The
CR distribution is nearly isotropic outside the MC, so if we denote
it at the MC edge as $f=f_{out}\left(p\right)$, then inside the MC,
instead of eq.(\ref{eq:finMCgen}), we write: $f=f_{in}\left(\mathcal{H}\right)=f_{out}\left[\sqrt{\mathcal{H}^{2}/c^{2}-m_{p}^{2}c^{2}}\right]$.
We dropped the second argument in eq.(\ref{eq:finMCgen}), $I$, because
it does not satisfy the isotropy condition. Returning to the variables
$x,p$, the solution inside the MC can be written as follows

\begin{equation}
f\left(p,\phi\right)=f_{out}\left[\sqrt{p^{2}+e^{2}\phi^{2}/c^{2}+2\left(e/c\right)\phi\sqrt{p^{2}+m_{p}^{2}c^{2}}}\right]\label{eq:fScatterFree}
\end{equation}

In the opposite case of frequent pitch-angle scattering, the largest
term of eq.(\ref{eq:KinEq}) is on its r.h.s.. For that reason, the
distribution must be nearly isotropic, $f\approx f_{0}\left(p,x\right)$.
Following a standard reduction to diffusive transport \citep{Jokipii66,M_PoP2015},
we eliminate the r.h.s. by averaging this equation over the pitch-angle:

\begin{equation}
\frac{\partial f_{0}}{\partial t}+\frac{Bv}{2p^{2}}\left.\frac{\partial}{\partial x}\right|_{\mathcal{H}}\frac{p^{2}}{B}\left\langle \left(1-\mu^{2}\right)\frac{\partial f}{\partial\mu}\right\rangle =0\label{eq:f0init}
\end{equation}
where we denoted

\[
f_{0}\equiv\frac{1}{2}\int_{-1}^{1}fd\mu\equiv\left\langle f\right\rangle 
\]
and 

\[
\left.\frac{\partial}{\partial x}\right|_{\mathcal{H}}\equiv\frac{\partial}{\partial x}-\frac{e}{v}\frac{\partial\phi}{\partial x}\frac{\partial}{\partial p}
\]
which is a derivative along the line of constant particle energy,
given by eq.(\ref{eq:Hamiltonian}). The averaged value $\left\langle \cdot\right\rangle $
in eq.(\ref{eq:f0init}) can be calculated perturbatively from eq.(\ref{eq:KinEq}),
considering the term on its r.h.s. as the leading one and ignoring
the $\partial f/\partial t$ on its l.h.s. This term is irrelevant
for the long time evolution equation for $f_{0}$ which we derive
here. Eq.(\ref{eq:f0init}) takes the then following form

\begin{equation}
\frac{\partial f_{0}}{\partial t}=\frac{vB}{p^{2}}\left.\frac{\partial}{\partial x}\right|_{\mathcal{H}}\frac{p^{2}\kappa}{vB}\left.\frac{\partial f_{0}}{\partial x}\right|_{\mathcal{H}}\label{eq:f0final}
\end{equation}
Here we have introduced a conventional diffusion coefficient

\[
\kappa=\frac{v^{2}}{4}\left\langle \frac{1-\mu^{2}}{D\left(\mu\right)}\right\rangle 
\]

It follows from eqs.(\ref{eq:fScatterFree}) and (\ref{eq:f0final})
that, assuming the CR distribution outside the MC to be $f=f_{out}\left(p\right),$
we have found it propagating into the MC along the levels of constant
$\mathcal{H}$ on the $x,p$ plane. This conclusion holds up for both
ballistic and diffusive propagation. In fact, as we argued earlier,
in sufficiently dense molecular clouds the CR propagate in part ballistically.
Namely, for particles with momenta

\begin{equation}
p_{1}<\left|p_{\parallel}\right|<p_{2},\label{eq:p12interval}
\end{equation}
there are no Alfven waves to resonate with, so that particles with
$\left|p_{\parallel}\right|>p_{1}$ propagate ballistically along
the lines $\mathcal{H}\left(p,x\right)=const$ on the $x,p$- plane.
Here the momenta $p_{1,2}$ are defined as follows

\selectlanguage{english}%
\begin{equation}
p_{1}=2V_{A}m_{p}\omega_{c}/\nu_{in},\;\;p_{2}=\frac{p_{1}}{4}\sqrt{\rho_{0}/\rho_{i}}>p_{1},\label{eq:p12}
\end{equation}
{where $V_{A}$ is the Alfven velocity, $\omega_{c}$
is the proton (nonrelativistic) gyrofrequency }$\omega_{c}=eB/m_{p}c$
{,
$\nu_{in}$ is the ion-neutral collision frequency, and $\rho_{0}/\rho_{i}\gg1$
is the ratio of the neutral to ion mass density. Particles with $\left|p_{\parallel}\right|<p_{1}$
propagate diffusively. Not surprisingly, the wave gap widens with
decreasing ionization rate $\rho_{i}/\rho_{0}$, eq.(\ref{eq:p12}). }

There are complications associated with the mixed propagation of CRs
in an MC. First, as the ballistic and diffusive propagation times
are different, a transient CR distribution inside the MC will develop
discontinuities at the boundaries in momentum space given by $p_{\|}=p_{1,2}$.
Moreover, if CRs enter the MC from one end and escape from the other,
as discussed above, a discontinuity at $p_{\|}=p_{1}$ must develop
even in a steady state, as argued in detail in \citep{MDS_APS12}.
To avoid these complexities, that are not inherent in the aspects
of MC electrodynamics we are concerned with here, we simplify the
treatment as follows. Assuming that the CR distribution function is
approximately the same on the two faces of MC, magnetically connected
through its interior, the problem becomes symmetric about the center
of MC, Fig.\ref{fig:ShockMCinteraction}. Under these circumstances,
the CR distribution which depends only on particle energy, eq.(\ref{eq:fScatterFree}),
is valid for both ballistic and diffusive propagation domains in the
momentum space. This simplification should not change the final result
concerning the positron injection from the MC into the shock acceleration
process significantly.

At a shock modified by the CR pressure, the spectrum is different
from that occurring in conventional shocks. The modified spectrum
can be represented by eq.(\ref{eq:DCasSol}) upstream ($x>0$) and
by $f\left(x,p\right)=f\left(0,p\right)$ downstream ($x\le0)$. For
a steady state, an upper cut-off momentum is imposed, but it does
not play a significant role, inasmuch we do not include the CR pressure
explicitely. The CR density integral, considered below, converges
at $p=\infty$, so we ignore the high-energy asymptotic of $f$ here.
For what follows, however, an important role plays the CR diffusion
coefficient $\kappa.$ In a subshock zone, where the CR intensity
is high and so is the level of self-driven turbulence, a Bohm diffusion
regime is likely to establish, $\kappa=vr_{g}/3$, where $v\left(p\right)$
and $r_{g}\left(p\right)$ are the CR speed and gyroradius. This regime
must change at the periphery of the shock precursor, but this region
is not important for the present treatment. It may also be seen that
eq.($\ref{eq:DCasSol}$), representing the solution of shock acceleration
problem, is not separable in $x$ and $p$, in the usual terms. An
important consequence of this property is a \emph{coordinate-dependent
low-energy cutoff}, at a momentum where $\Psi\left(x\right)\sim\kappa\left(p\right)$. 

The CR number density in the precursor can thus be written as follows

\begin{equation}
N_{{\rm CR}}\left(x\right)=4\pi\int_{0}^{\infty}p^{2}F\left(p\right)e^{-q_{b}\left(p\right)\Psi\left(x\right)/3\kappa\left(p\right)}dp\label{eq:CRdensPrec}
\end{equation}
Generally speaking, the lower integration limit should be equal to
an injection momentum, $p_{{\rm inj}},$ at which the solution in
eq.(\ref{eq:DCasSol}) should be matched with the thermal distribution.
The matching can be performed with some overlapping between the above
solution and an intermediate asymptotic solution that, on the lower
energy end, smoothly transitions into the thermal distribution \citep{m98}.
However, as we primarily interested in the upstream spectrum, for
$\Psi\left(x\right)>\kappa\left(p_{{\rm inj}}\right),$ we replaced
$p_{{\rm inj}}$ by zero in eq.(\ref{eq:CRdensPrec}). 

We need to know the CR density inside the MC, while the last expression
provides this quantity at a distance $x$ from the subshock and can
only be considered as boundary condition for the CR distribution inside
the MC. Therefore, we evaluate $N_{CR}$ in eq.(\ref{eq:CRdensPrec})
as follows. First, normalizing the proton momentum to $m_{p}c$, we
specify the diffusion coefficient

\begin{equation}
\kappa=\frac{\kappa_{0}p^{2}}{\sqrt{1+p^{2}}}\label{eq:KappaBohm}
\end{equation}
where $\kappa_{0}\sim c^{2}/\omega_{c}$ is the reference diffusivity
of a mildly relativistic proton ($\omega_{c}$ denotes the nonrelativistic
cyclotron frequency). Next, we substitute this $\kappa$ into eq.(\ref{eq:CRdensPrec}),
bearing in mind that the main contribution to the CR density comes
from mildly relativistic protons. In this range, their spectrum is
close to $p^{-4}.$ Thus, we find $N_{CR}\left(x\right)\propto1/\sqrt{\Psi}.$
In fact, the contribution of higher energy protons, where the spectrum
hardens to $p^{-7/2}$ does not change this result significantly.
Indeed, at ultra-relativistic momenta, $\kappa$ also changes its
scaling to $\kappa\propto p$, and the momentum differential under
the integral in eq.(\ref{eq:CRdensPrec}) can be replaced by $d$$\kappa^{-1/2}$
in both cases. Thus, the coordinate dependence $N_{CR}\left(x\right)\propto1/\sqrt{\Psi}$
holds up. Using eq.(\ref{eq:FlowPotDef}), after some obvious notation
changes, this dependence can be transformed to eq.(\ref{eq:NcrOfx}).

Now we turn to the CR distribution inside an MC, provided by eq.(\ref{eq:fScatterFree}),
given the electrostatic potential and CR momentum distribution at
the MC boundary, discussed above. For the equilibrium solution to
be valid, the boundary condition should change slower than the CR
propagation time. This is certainly true for the ballistic CR propagation,
but it is only marginally acceptable for the diffusive regime unless
the CR diffusivity inside the MC is larger than that outside. Note
that this condition is met if there is a strong collisional damping
of Alfven waves that confine CRs inside the MC. We assume it to be
valid here, for simplicity. The goal is to understand, what is the
MC response to the external charge brought in by the penetrating CRs.

Let us consider a magnetic field line threading an MC and intersecting
its surface at the points $x=\pm a$. We count the $x$ coordinate
from the center of the MC on the field line. On assuming the CR distribution
to be the same at the end points, $\pm a$, the problem of MC charging
by the CRs and their neutralization by return currents becomes symmetric
in $x$. Hence, we can fix the boundary condition for the induced
electrostatic potential as $\partial\phi/\partial x=0$ at $x=0$
and $\phi=0$ at $x=\pm a$. In equilibrium, the CRs do not escape
from the MC but merely interchange with CRs outside. Consequently,
they evenly populate the lines of constant $\mathcal{H}\left(x,p\right)=const$,
eq.(\ref{eq:Hamiltonian}), in the particle phase space along the
field lines. So, we can use eq.(\ref{eq:fScatterFree}) for the number
density of CR inside the MC. Regardless the propagation regime, ballistic
or diffusive, the CR number density inside the MC can be written down
as a function of $\phi$ 

\begin{equation}
N_{CR}\left(\phi\right)=4\pi\int p^{2}f_{out}\left(\sqrt{p^{2}+e^{2}\phi^{2}/c^{2}+2\left(e/c\right)\phi\sqrt{p^{2}+m^{2}c^{2}}}\right)dp\label{eq:NcrOfFi}
\end{equation}
Here $f_{out}\left(p\right)$ is the CR distribution at the MC boundary,
where $\phi=0$.

\section{Electrodynamics Inside MC\label{sec:Appendix:-electrodynamics-inside}}

\begin{figure}
\includegraphics[scale=0.6]{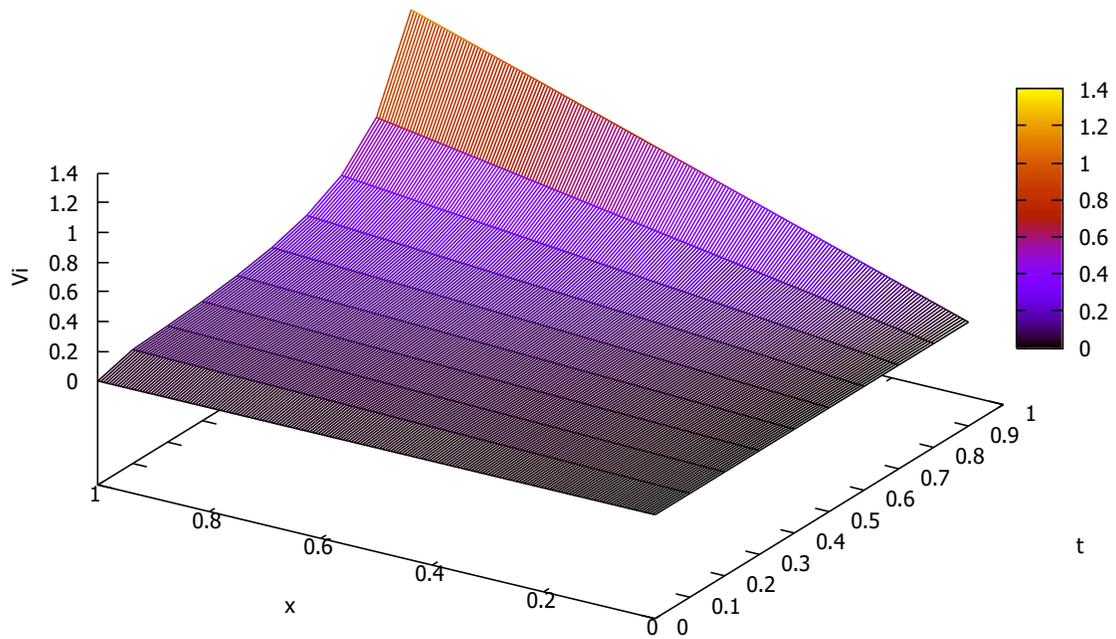}

\caption{Time evolution of the ion velocity profile, starting from $V_{i}\left(x,0\right)\equiv0$,
as described by eqs.(\ref{eq:dVdtResc}-\ref{eq:dndtResc}). The CR
source term is prescribed according to eq.(\ref{eq:NcrOft}), $n_{CR}\propto1/\left(t_{0}-t\right).$
The ion density remains constant in $x$ at all times. Here $x=0$
corresponds to the mid point of the MC, while $x=1$ \textendash{}
to its edge.\label{fig:Time-evol of Vi}}

\end{figure}

First we demonstrate that the reduction of the PDE system of eqs.(\ref{eq:dVdtResc}-\ref{eq:dndtResc}),
to ODE system of eqs.(\ref{eq:dpsidt}-\ref{eq:dnidt}), via the representation
given by eqs.(\ref{eq:niOft}-\ref{eq:Voft}) (homogeneous deformation
flow), is a robust attractor of the PDE solution. We demonstrate this
by integrating the PDE system directly. The result is illustrated
in Fig.\ref{fig:Time-evol of Vi} which shows the profile of $V_{i}\left(t,x\right)$.
Starting from the rest, the flow adheres to a perfectly linear profile
in $x$ at all times. The ion density $n_{i}\left(t,x\right)$ remains
constant in $x$, as it should to.

Now we turn to eqs.(\ref{eq:dpsidt}-\ref{eq:dnidt}) that can be
further simplified by assuming $n_{CR}\ll n_{i}$ and denoting $F=\nu_{e}n_{CR}$
:

\begin{eqnarray}
\frac{\partial\psi}{\partial t} & = & -\psi\left(\psi+1\right)+F\psi+\frac{\partial F}{\partial t}\label{eq:PhiEq}\\
\frac{\partial n_{i}}{\partial t} & = & -\psi n_{i}\label{eq:niIndepEq}
\end{eqnarray}
The function $F\left(t\right)$ is derived here from $n_{CR}\left(t\right)$,
given in eq.(\ref{eq:NcrOft})

\[
F=\frac{\alpha}{t_{0}-t}
\]
The dimensionless parameters $\alpha$ and $t_{0}$ play an important
role in the analysis and are defined as follows (see eqs.{[}\ref{eq:NcrOfx},\ref{eq:NcrOft}{]})

\begin{equation}
\alpha=\frac{m_{e}}{m_{i}}\frac{a}{u_{1}}n_{CR}^{0}\nu_{ei}^{\prime},\;\;\;t_{0}=\nu_{in}a/u_{1}\label{eq:alphaAndt0}
\end{equation}
Recall that $x_{0}$ denotes an MC's closest approach to the subshock.
Clearly, $x_{0}\sim a,$ the size of MC, so we simply substituted
$a$ for $x_{0}$ in the above parameters. The first factor entering
the parameter $\alpha$ is small, $\lesssim10^{-3}$, but the remaining
combination of parameters has a meaning of the number of $e-i$ collisions
inside the cloud during its shock crossing, diminished by the factor
$n_{CR}^{0}/n_{i}$. Overall, $\alpha$ may become quite large for
big MCs.

Our next goal is to understand how the ion velocity $\psi$ grows
with time while the MC approaches the subshock from $+\infty$. The
source function $F\left(t\right)\propto n_{CR}\left(t\right)$ and
we assume that time is changing from $t=-\infty$ to $t=0.$ At this
last moment, the MC starts crossing the subshock and positrons generated
in its interior, are already largely expelled by the electric field.
This final value of the electric field and a closely related value
of ion outflow velocity comprise the main subject of the analysis
below. 

Eq.(\ref{eq:PhiEq}) is independent of eq.(\ref{eq:niIndepEq}) and
it is easy to guess its particular exact solution

\begin{equation}
\psi_{1}\left(t\right)=F-1.\label{eq:Psi1}
\end{equation}
This solution is, however, unphysical since it does not behave properly
at $t=-\infty$. Nonetheless, we will use it to find the solution
with the proper behavior at $t=-\infty$, that is $\psi\to0$, as
$t\to-\infty$. For, we linearise eq.(\ref{eq:PhiEq}) using a conventional
substitution for Riccati equations

\begin{equation}
\psi=\frac{1}{w}\frac{\partial w}{\partial t}+\frac{1}{2}\left(F-1\right)\label{eq:FhiDefin}
\end{equation}
and obtain the following equation for $w$ 
\begin{equation}
\frac{\partial^{2}w}{\partial t^{2}}-Qw=0\label{eq:wLinEq}
\end{equation}
where we denoted

\[
Q=\frac{1}{4}\left[\left(F-1\right)^{2}+2\frac{\partial F}{\partial t}\right]
\]
Eq.(\ref{eq:wLinEq}) can be solved in terms of Whittaker functions
but, as we have already obtained one particular solution, it is easier
to find the required solution directly. On denoting $w_{1}\left(t\right)$
the solution of eq.(\ref{eq:wLinEq}), that corresponds to $\psi_{1}$
in eq.(\ref{eq:Psi1}), we find from eq.(\ref{eq:FhiDefin}) with
$\psi=\psi_{1}:$

\[
w_{1}=e^{\tau/2}\tau^{-\alpha/2}
\]
where we used the notation $\tau\equiv t_{0}-t$. Now we can find
the second linearly independent solution to eq.(\ref{eq:wLinEq}),
$w_{2}$, as follows

\[
w_{2}=Cw_{1}\int\frac{d\tau}{w_{1}^{2}}
\]
where $C$ is an arbitrary constant which does not play any role given
the relation between $w$ and $\psi$ in eq.(\ref{eq:FhiDefin}).
Returning to the original variable $\psi$ by substituting the last
expression into eq.(\ref{eq:FhiDefin}), we find the required solution
of eq.(\ref{eq:PhiEq}):
\begin{equation}
\psi_{2}\left(\tau,\alpha\right)=\frac{\alpha}{\tau}-1+\frac{\tau^{\alpha}e^{-\tau}}{\int_{\tau}^{\infty}t^{\alpha}e^{-t}dt}\label{eq:psi2App}
\end{equation}
We rewrite this result in an equivalent form, dropping the subscript
at $\psi$ in eq.(\ref{eq:psiOftauandalpha}).

\bibliographystyle{prsty}
\bibliography{}

\clearpage

\end{document}